\documentclass{jfm}
\usepackage{graphicx}
\usepackage{epstopdf, epsfig}
\usepackage{amsmath}
\usepackage{amssymb}
\usepackage{bm}
\usepackage{color}
\usepackage{amsbsy}
\usepackage{tabularx}
\usepackage{enumerate}
\usepackage{url}

\DeclareMathOperator{\grad}{grad}
\DeclareMathOperator{\divr}{div}
\DeclareMathOperator{\spn}{span}

\shorttitle{Koopman analysis of turbulent convection}
\shortauthor{D. Giannakis, A. Kolchinskaya, D. Krasnov and J. Schumacher}

\title{Koopman analysis of the long-term evolution in a turbulent convection cell}

\author{Dimitrios Giannakis\aff{1},
            Anastasiya Kolchinskaya\aff{2},
            Dmitry Krasnov\aff{2}
 \and   J\"org Schumacher\aff{2}
  \corresp{\email{joerg.schumacher@tu-ilmenau.de}} }

\affiliation{\aff{1}Center for Atmosphere Ocean Science, Courant Institute of Mathematical Sciences, New York University, New York, NY 10012, USA
\aff{2} Institut f\"ur Thermo- und Fluiddynamik, Postfach 100565, Technische Universit\"at Ilmenau, D-98684 Ilmenau, Germany}

\begin{document}

\maketitle

\begin{abstract}
We analyse the long-time evolution of the three-dimensional flow in a closed cubic turbulent Rayleigh-B\'{e}nard convection 
cell via a Koopman eigenfunction analysis. A data-driven basis derived from diffusion kernels known in machine learning is 
employed here to represent a regularized generator of the unitary Koopman group in the sense of a Galerkin approximation. 
The resulting Koopman eigenfunctions can be grouped into subsets in accordance with the discrete symmetries in a cubic box. 
In particular, a projection of the velocity field onto the first group of eigenfunctions reveals the four stable large-scale circulation 
(LSC) states in the convection cell. We recapture the preferential circulation rolls in diagonal corners and the short-term switching 
through roll states parallel to the side faces which have also been seen in other simulations and experiments. The diagonal macroscopic 
flow states can last as long as a thousand convective free-fall time units. In addition, we find that specific pairs of Koopman eigenfunctions 
in the secondary subset obey enhanced oscillatory fluctuations for particular stable diagonal states of the LSC. The corresponding
velocity field structures, such as corner vortices and swirls in the midplane, are also discussed via spatiotemporal reconstructions. 
\end{abstract}


\section{Introduction}
The global turbulent transport of heat and momentum in turbulent Rayleigh-B\'{e}nard convection depends sensitively on the flow geometry, 
particularly in low aspect ratio domains as shown for example in simulations by \cite{Bailon2010}, \cite{Kaczorowski2013} and \cite{Chong2016}. 
In addition, the global transport properties depend on the specific shape of the cell or the container \citep{Daya2001,Song2014}. The reason 
for this behavior is a complex-shaped large-scale circulation (LSC) flow, formed in closed convection volumes. The LSC fills the whole 
cell, and evolves on several characteristic timescales simultaneously. One timescale is the typical turnover time of a fluid element in an LSC roll.
The LSC roll itself can change its mean orientation on a further characteristic timescale. In addition, this circulation shows frequently an internal torsion that 
results in a sloshing motion, especially in cylindrical geometries \citep{Zhou2009}. 

While the LSC roll drifts slowly in the azimuthal direction in a 
cylinder \citep{ShiEtAl12} -- the only remaining statistically homogeneous direction -- it gets locked in diagonal corners or parallel to the side walls in cubic and rectangular containers. In a cubic container, the LSC switches rather rapidly from one into another of the eight possible macroscopic flow states \citep{Foroozani2014,Bai2016,Foroozani2017}. The dynamics is then determined by discrete symmetries such as rotations by 90, 180, 270, and 360 degrees. Also, for cubic or rectangular geometries statistical homogeneity is completely absent, and the turbulent flow is fully inhomogeneous. Different numbers of LSC rolls are possible when the container is rectangular. In this geometry, long-term large-eddy simulations exhibit switching between these states on a very long timescale of order $ 10^3$ convective time units  \citep{Podvin2012}. However, such long transients of the LSC are typically not accessible with direct numerical simulation (DNS),  particularly when the Rayleigh number $ Ra $  of the convection flow exceeds $\sim 10^5$. This calls for a reduction of the degrees of freedom of the system.

Models with different levels of reduction have been developed to understand the long transients of the LSC dynamics,  among them a nonlinear oscillator 
model for a cylindrical cell \citep{Brown2006} and a three-mode LSC model based on a Proper Orthogonal Decomposition 
(POD) in a square convection cell \citep{Podvin2015}. POD-type extraction of the most energetic degrees of freedom has also been used in the three-dimensional 
case to determine the contribution of the primary and secondary POD modes to the global turbulent heat transfer \citep{Bailon2010}.  

A drawback of POD analysis is that it is prone to mixing multiple timescales in the recovered modes, compounding their physical interpretation. In contrast, Dynamic Mode Decomposition (DMD) \citep{SchmidSesterhenn08,Schmid10}
generally has higher skill in separating the different timescales in the dynamics 
as individual modes, as recently demonstrated for rotating convection in a cylindrical cell by \cite{Horn2017}.
This capability of DMD stems from its connections with Koopman operators of dynamical systems \citep{RowleyEtAl09,Mezic13}. The latter, are the natural linear operators governing the evolution of observables under the (potentially nonlinear) dynamics, and have many useful spectral properties for data-driven applications \citep{MezicBanaszuk04,Mezic05}. For instance, in ergodic systems, the Koopman operators are unitary, and their eigenfunctions are periodic observables (even though the dynamical flow may be non-periodic) evolving at characteristic timescales determined by their corresponding eigenvalues.

Despite their attractive theoretical properties, a major challenge with calculating Koopman eigenvalues and eigenfunctions is their 
approximation from high-dimensional data, such as velocity or temperature field snapshots of fully turbulent convection flows. 
In DMD, that problem is reduced to approximating the eigenvalues and the projections of the data onto Koopman eigenfunctions, 
as opposed to the eigenfunctions themselves which are generally harder to compute. These projections are spatial patterns called Koopman modes \citep{Mezic05}.  
Convergence results for the Koopman eigenvalues and 
eigenfunctions are available in more recently developed methods such as  extended DMD \citep[][]{WilliamsEtAl15} and Hankel matrix DMD \citep[][]{TuEtAl14,ArbabiMezic16,BruntonEtAl17}. These methods generally operate directly in snapshot spaces or in spaces of sequences of snapshots 
in the case of Hankel DMD, with an associated high computational cost, and can sometimes yield spurious unstable modes. \citet{WilliamsEtAl15b} 
have developed a kernel-based variant of DMD  with an improved computational cost in high-dimensional data spaces, though to our knowledge, 
no convergence results have been reported for this class of algorithms.  

In general, the unitary Koopman group and its generator are approximated with respect to a basis or dictionary for the Hilbert space of observables under study. These operators usually have a continuous spectrum and must be regularized (e.g., by means of diffusion) to obtain a discrete spectrum. Recently, it was shown that kernel integral operators employed in machine learning can provide data-driven orthonormal bases for approximating Koopman operators that can be efficiently computed from high-dimensional data, while providing guarantees in the asymptotic limit of large data \citep{GiannakisEtAl15,Giannakis17,DasGiannakis17}. Thus, these methods overcome some of the obstacles in DMD and related algorithms. Moreover, the same class of kernel operators naturally leads to data-driven diffusion operators for regularization and an associated notion of smoothness of observables, e.g.\ by a Dirichlet energy functional.

Here, we construct our data-driven basis using the kernels introduced in so-called nonlinear Laplacian spectral analysis (NLSA) algorithms \citep{GiannakisMajda11c,GiannakisMajda12a,GiannakisMajda13} and in independent work by \citet{BerryEtAl13}. This class of kernels operates on delay-coordinate mapped data as in singular spectrum analysis (SSA)  \citep{BroomheadKing86,VautardGhil89,GhilEtAl02} and Hankel matrix DMD, and further employs a Markov normalization procedure introduced in the diffusion maps algorithm for manifold learning \citep{CoifmanLafon06}. As shown by \citet{Giannakis17} and \citet{DasGiannakis17}, the kernel eigenfunctions obtained via NLSA provide an efficient basis for Galerkin approximation of Koopman operators since, as the number of delays increases, they span finite-dimensional invariant subspaces associated with the point spectrum of the Koopman operator. NLSA has previously been used to study heat transfer in two-dimensional Rayleigh-B\'enard convection in a large-aspect-ratio periodic domain \citep{BrenowitzEtAl16}. 

In the present work, we apply these techniques to a fully turbulent three-dimensional convection flow in a closed cubic cell. In particular, we analyse a dataset from a long-term DNS extending over 10,000 free-fall time units, containing 10,000 equidistantly written fully resolved velocity field snapshots at a grid resolution of $128^3$ grid points for a Rayleigh number of $Ra=10^7$ and a Prandtl number of $Pr=0.7$. Using time-ordered velocity field snapshots from this dataset, we identify LSC patterns of the flow through Koopman eigenfunctions. Our analysis will reveal that the dominant LSC configurations of this convection flow, as well as their associated transition timescales, can be described in terms of a small set of Koopman eigenfunctions, thus establishing a connection between the evolution of the LSC and the intrinsic spectral properties of the Boussinesq system in this regime. 

Our present analysis is restricted to the flow behavior 
as opposed to a joint analysis of velocity and temperature fields (e.g., as in \citet{Bailon2010}), 
and focuses on the changes of the large-scale flow in the cell. We will show that 
the dynamics at the largest scales is obtained from the subset of Koopman eigenfunctions with the lowest Dirichlet energy, which evolve on timescales of the order of $10^3$ free-fall time units, characteristic of the LSC. The first three such eigenfunctions describe the long-term LSC switching between different symmetry states associated with the cubic flow geometry. The next group of eigenfunction pairs represents secondary structures such as corner vortices. Particular eigenfunctions in this group are found to exhibit enhanced fluctuations when the LSC flow is in one of the four stable diagonal LSC states; a subset of the eight possible macrostates mentioned above. 

The manuscript is organized as follows. Section~\ref{secBoussinesq} describes the simulation model, and lists some essential properties of the flow.
Section~\ref{secKoopman} introduces the general ideas of the Koopman eigenfunction analysis, and describes the associated Galerkin method in the data-driven basis acquired from NLSA, as well as the procedure for spatiotemporal reconstruction based on Koopman eigenfunctions. This theoretical section is followed by section~\ref{secResults}, 
which discusses in detail the first two sets of Koopman eigenfunctions, namely three primary Koopman eigenfunctions and four pairs of secondary eigenfunctions. The section also includes an additional ensemble analysis to shed more light on the long-time behaviour of the large-scale flow states in the cell. We conclude the paper in section~\ref{secSummary} with a summary and outlook. A discussion on numerical implementation of our Koopman eigenfunction framework is included in appendix A.

\section{\label{secBoussinesq} Direct numerical simulation data}
We solve the three-dimensional equations of motion in a cubic domain $ \Omega = [0,1]\times [-0.5,0.5]\times [-0.5,0.5]$ with Cartesian coordinates
in the Boussinesq approximation 
\citep{Chilla2012}. The equations are made dimensionless by using the height of the cell $H$, the free-fall velocity $U_f=\sqrt{g \alpha \Delta T H}$, and 
the imposed temperature difference $\Delta T$ as characteristic length, velocity, and temperature scales, respectively. The equations contain three 
control parameters: the Rayleigh number $Ra$, the Prandtl number $Pr$, and the aspect ratio $\Gamma=L/H=1$ with the side length $L$. The 
dimensionless equations are given by
%-------------------------------------------------------------------------------
\begin{subequations}
  \label{eqBoussinesq}
\begin{align}
\label{ceq}
{\bm \nabla}\cdot {\bm u}&=0\,,\\
\label{nseq}
\frac{\partial{\bm u}}{\partial  t}+({\bm u}\cdot{\bm\nabla}){\bm u}
&=-{\bm \nabla}  p+\sqrt{\frac{Pr}{Ra}} {\bm \nabla}^2{\bm u}+  T {\bm e}_z\,,\\
\frac{\partial  T}{\partial  t}+( {\bm u}\cdot {\bm \nabla})  T
&=\frac{1}{\sqrt{Ra Pr}} {\bm \nabla}^2  T\,,
\label{pseq}
\end{align}
\end{subequations}
%-------------------------------------------------------------------------------
with 
%-------------------------------------------------------------------------------
\begin{displaymath}
Ra=\frac{g\alpha\Delta T H^3}{\nu\kappa}=10^7\,,\;\;\;\;\;\;\;\;Pr=\frac{\nu}{\kappa}=0.7\,.
\end{displaymath}
%-------------------------------------------------------------------------------
Here, the variable $g$ stands for the  acceleration due to gravity, $\alpha$ is the thermal expansion coefficient, $\nu$ is the kinematic viscosity, and 
$\kappa$ is the thermal diffusivity. No-slip boundary conditions for the fluid (${\bm u}=0$)  are applied at all walls. The side walls are thermally insulated 
($\partial T/\partial {\bm n}=0$), and the top and bottom plates are held at constant dimensionless temperatures $T=0$ and 1, respectively. Times 
are measured in free-fall time units $T_f=H/U_f$. 

We will use the notation $ x = ( \bm u, T ) $ to represent the state of the Boussinesq system associated with a velocity field $ \bm{ u } \in 
L^2( \Omega; \mathbb{ R}^3 ) $ and the coupled temperature field $ T \in L^2( \Omega ) $. The corresponding space of convection states
will be denoted by $ X = L^2( \Omega; \mathbb{ R }^3 ) \oplus L^2( \Omega ) $. We also let 
%-------------------------------------------------------------------------------
\begin{equation}
F : X \mapsto L^2( \Omega; \mathbb{ R }^3 )
\label{proj}
\end{equation}
%-------------------------------------------------------------------------------
be the projection map mapping states $ x = ( \bm u, T ) $ in $ X $ to the corresponding velocity fields, viz.\ $ F( x ) = \bm{ u } $. For our purposes, $F$ will act as the observation function, mapping states of the Boussinesq system to the velocity field data that we analyze.

%------------------------------------------------------------------------------------------
\begin{figure}
\begin{center}
\includegraphics[width=5.5in,angle=0]{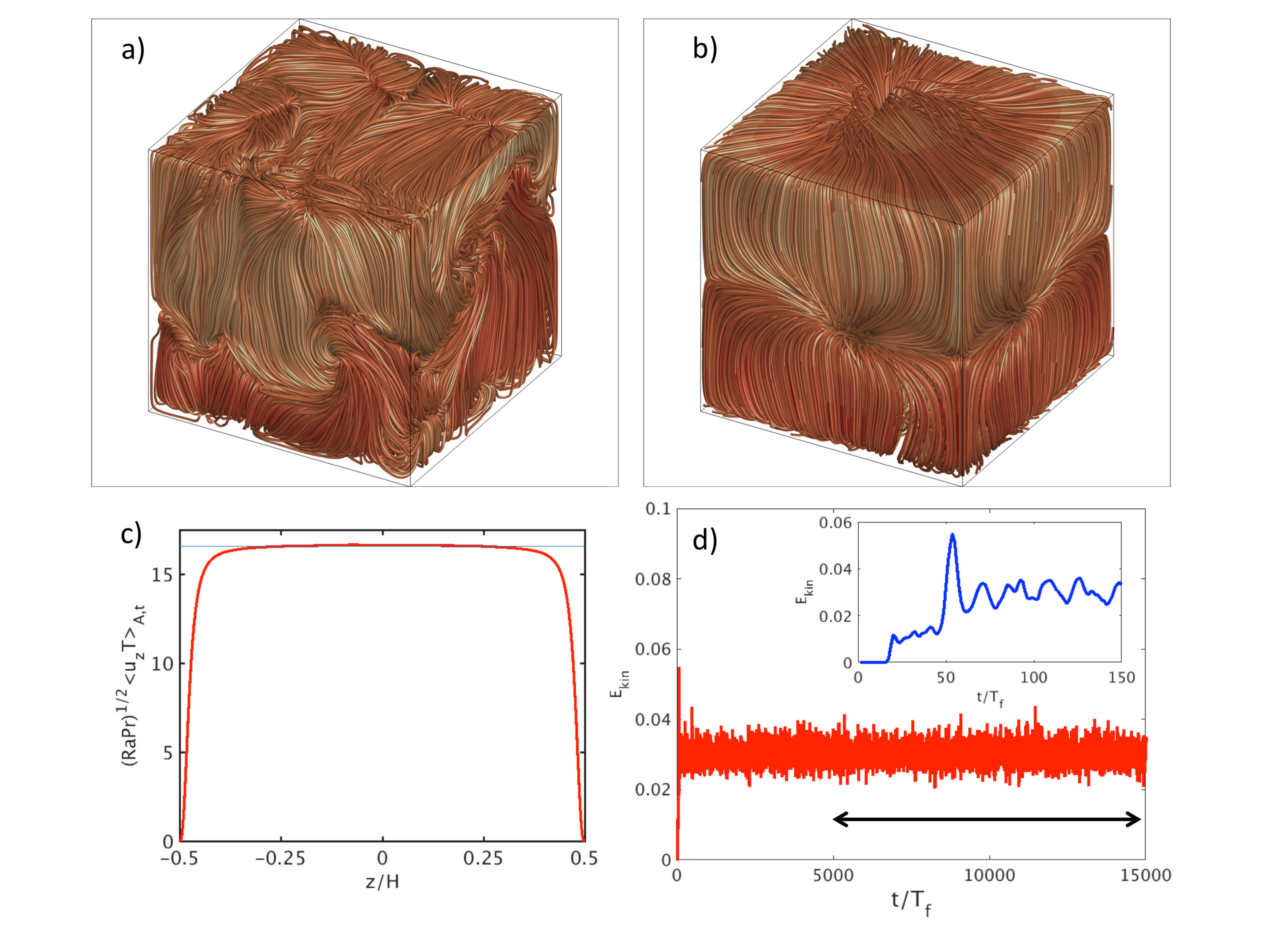}
\caption{Turbulent convection flow in a closed cubic cell. (a) Instantaneous snapshot of velocity field lines. (b) Field line plot of 
time-averaged velocity field. Time average is taken over 10,000 $T_f$. (c) Convective heat transfer as a function of the vertical coordinate 
$z$. The horizontal line shows the global Nusselt number $Nu$ from~\eqref{NuV} for reference. (d) Turbulent kinetic energy versus 
time. The double-headed arrow which spans from $t/T_f=$ 5,000 to 15,000 indicates the 10,000 free-fall time units for our  analysis. 
The inset illustrates the initial relaxation into a fully developed turbulent state.}
\label{Fig1}
\end{center}
\end{figure}
%-------------------------------

In response to the input parameters $Ra$, $Pr$, and $\Gamma$,  turbulent heat and momentum fluxes 
are established. The turbulent heat transport is determined by the Nusselt number  
\begin{equation}
  \label{NuV}
Nu=1+\sqrt{Ra Pr}\langle  u_z  T\rangle_{\Omega,t}\,.
\end{equation} 
For the given input parameters, we find $Nu=16.57 \pm 0.02$ which is in agreement with other DNS studies at these Rayleigh and Prandtl numbers \citep[e.g.,][]{Bailon2010}. The turbulent  momentum transfer is given by the Reynolds number 
\begin{displaymath}
Re= \sqrt{\langle u_i^2\rangle_{\Omega,t}Ra/Pr}\,, 
\end{displaymath}
which amounts to $Re=654 \pm 1$ and is comparable with the study of \cite{Scheel2014}.  In both definitions $\langle\cdot\rangle_{\Omega,t}$ 
stands for a combined volume and time average. The equations are numerically solved by a second-order finite difference method by 
\cite{Krasnov2011} on a non-uniform mesh which gets finer towards the walls. In this configuration, 16 horizontal grid planes are found inside the 
thermal boundary layer with a nominal thickness $\delta_T=1/(2 Nu)\approx 0.03$. 

Figure \ref{Fig1} displays instantaneous (figure~\ref{Fig1}(a)) and time-averaged (figure~\ref{Fig1}(b)) field line plots of the velocity field.  In the latter case, time averaging (performed over the full analysis interval) reveals the LSC as a diagonally oriented bundle of streamlines originating from an impingement point in the back corner. In the lower right corner of the same plot one observes a recirculation 
vortex very similar to what has been shown by \citet{Foroozani2017}. Note that due to the off-center location of the impingement point the time-averaged flow is not invariant under the full discrete symmetry group associated with the cubical flow domain. Figure~\ref{Fig1}(c) displays the plane-averaged profile of the convective 
heat flux, $\langle u_z T(z)\rangle_{A,t}$, with $ A $ standing for horizontal area averaging. Away from the boundary layers, this quantity is essentially identical to  
the globally determined Nusselt number via~\eqref{NuV}. Figure~\ref{Fig1}(d) displays 
the kinetic energy in the cube as a function of time. Based on these results, we have taken 5,000 $T_f$ to relax the system into a statistically stationary turbulent state before the Koopman analysis begins. In separate calculations with smaller relaxation times, we have observed a very 
slow transient, visible only in the phase portraits of the leading Koopman eigenfunctions, which made this long relaxation necessary.
Interestingly, this transient is not detectable when standard statistical moments of the Eulerian turbulent fields are inspected, such as root mean square 
velocities or turbulent kinetic energy.

\section{\label{secKoopman}Koopman operator formalism}

In this section, we summarize the mathematical framework underlying our data analysis. Our discussion is self-contained, but does not cover certain technical aspects of this formalism. The reader interested in these topics is referred to the listed references.      

\subsection{\label{secDefKoop}Definition of the Koopman operator and its infinitesimal generator}
Let  $ x_{-Q+1}, x_{-Q+2}, \ldots, \ldots, x_{N-1} $ be a sequence of states $ x_n = ( \bm u_n, T_n ) \in X $ of the Boussinesq system from section~\ref{secBoussinesq}, sampled at times $ t_n = n \tau$, where $\tau$ is 
a fixed sampling interval, and $ Q, N $ are positive integers. In what follows, $ x_{-Q+1}  $ will be the first state sampled after the initial equilibration interval in our simulations. For the DNS data analysis we will take $N=10,000$, $Q=30$, and $ \tau=1$ (the latter, in units of the free-fall time $T_f$).  
Note that we assign the potentially negative timestamp 
$ t_{-Q+1} $ to the first sample in our dataset in anticipation of the fact that we will be performing delay-coordinate maps with $ Q $ delays---this ``uses up'' the 
first $ Q - 1 $ samples, so we will be left with $ N $ available samples for analysis starting at time $ t_0 = 0 $.

Formally, we assume that our Boussinesq system admits a compact invariant set $ A \subset X $ (e.g., an attractor) with a compact, forward-invariant 
neighborhood $ \mathcal{ U } $ in which $ x_0 $ lies. We also assume that  $x_0$ lies in the basin of a physical ergodic  probability measure 
$ \mu $ supported on $ A $ \citep[][]{Young02}. Under this assumption, given any continuous function 
$ f : X \mapsto \mathbb{ C } $ (also referred to as observable), its time average $ \bar f_N $ along the orbit $ \{ x_n \}_{n=0}^{N-1} $ converges to its expectation value with respect to the invariant measure, which translates here to
\begin{displaymath}
\lim_{N\to\infty} \bar{f}_N=\lim_{N\to\infty} \frac{1}{N} \sum_{n=0}^{N-1} f(x_n)=\int_A f \, d\mu \,.
\end{displaymath}
These  assumptions, which cannot be rigorously justified but are tacitly made in many long-time statistical analyses of such systems, 
are sufficient to ensure that the results of our data-driven algorithms converge in the limit of large data, $ N \to \infty $ \citep[][]{DasGiannakis17}.
 
\begin{figure}
\begin{center}
\includegraphics[width=4in,angle=0]{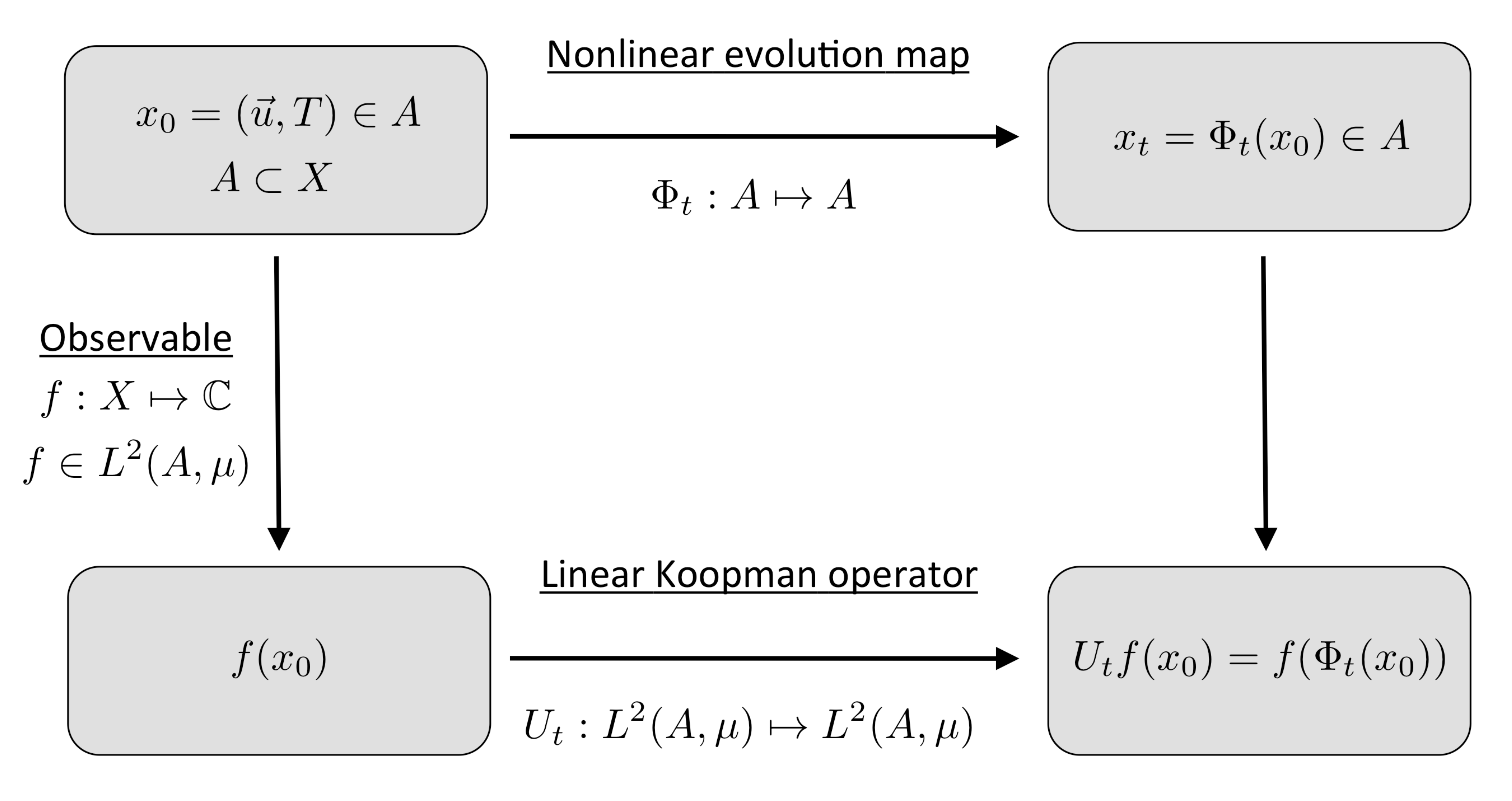}
\caption{Connection between nonlinear evolution map and the linear Koopman operator in a commutative diagram. }
\label{Diagram}
\end{center}
\end{figure}

The dynamics of our turbulent convection flow in state space can be characterized through a nonlinear evolution map 
$ \Phi_t : X \mapsto X $, $ t \in \mathbb{ R }$, such that $ \Phi_t(x) $ is the state reached at time $ t $ under dynamical evolution by the 
Boussinesq system~\eqref{eqBoussinesq} with initial data $ x \in X$. Moreover, associated with $ \Phi_ t $ is a unitary group of Koopman operators 
$ U_t : L^2( A, \mu ) \mapsto L^2( A, \mu ) $, acting on observables in the Hilbert space $ L^2(A,\mu) $ of square-integrable (with respect 
to $ \mu$), complex-valued functions on the attractor \citep[][]{EisnerEtAl15}. These linear operators act on $f \in L^2(A,\mu) $ via composition 
with the flow map, i.e., $ U_t f = f \circ \Phi_t $, and for all $ t \in  \mathbb{ R } $ and $ f \in L^2(A, \mu ) $ the map $ t \mapsto U_t f $ is 
continuous. As with all such continuous one-parameter unitary groups,  $U_t$ is fully determined by its infinitesimal generator, $ V: D( V ) 
\mapsto L^2( A, \mu ) $. This operator is an unbounded, skew-adjoint operator with a dense domain $ D( V ) \subset L^2(A, \mu ) $, 
acting on observables $f \in D(V)$ via
\begin{displaymath} 
 V f = \lim_{t \to 0} \frac{1}{t} ( U_t f - f )\,.
\end{displaymath}  
The  Koopman operator at any $ t \in \mathbb{R} $ can be recovered via exponentiation of $ V$, viz.\ $ U_t = \exp(t V )$. Figure \ref{Diagram} summarizes this 
connection between the linear Koopman operator and the nonlinear evolution (or flow) map. 
%------------------------------------------------------------------------------------------
%-------------------------------

The action of an ergodic dynamical system on observables can be characterized in terms of the spectral properties of Koopman operators 
$U_t$ \citep[][]{Mezic05}, or, equivalently, the related transfer operators \citep[][]{DellnitzJunge99} which are adjoint to $ U_t$. Another 
equivalent characterization, which we will adopt here, is through eigenfunctions of the generator $ V $. In particular, a distinguished class of 
observables are Koopman eigenfunctions, satisfying the eigenvalue problem $ V \psi = \lambda \psi $ for some $ \psi \in D( V ) $ and 
$ \lambda \in \mathbb{ C } $. For measure-preserving dynamical systems (including the ergodic systems of interest here), every nonzero eigenvalue $ \lambda = i \omega $ of $ V $ is purely 
imaginary, and captures an intrinsic frequency $ \omega \in \mathbb{ R } $ of the system. Moreover, the corresponding eigenfunctions 
$ \psi $ evolve periodically under the dynamics at the frequencies associated with the eigenvalues; that is, $ U_t \psi = \exp(i \omega t) \psi $. Because $ V f^* = ( V f )^* $ for any $ f \in D( V ) $, the Koopman eigenvalues and eigenfunctions occur as complex conjugate pairs; i.e., if $ \psi $ is an eigenfunction at eigenvalue $ \lambda $, then $ \psi^* $ is an eigenfunction at eigenvalue $ \lambda^* = - \lambda $. For ergodic systems, all eigenvalues of $ V $ are simple. Throughout this paper, we will assume that all Koopman eigenfunctions are continuous (to our knowledge, there are no known continuous-time ergodic dynamical systems with discontinuous Koopman eigenfunctions).  

Evaluated along the convection flow trajectory in state space given by $ x_0, x_1, \ldots $, each Koopman eigenfunction induces a time series 
$ \psi( x_0 ), \psi( x_1 ), \ldots $ that can be thought of as an analog of a temporal (chronos) mode in a POD \citep[][]{AubryEtAl91}, but with 
potentially higher dynamical significance. For our purposes, of particular interest are eigenfunctions at small corresponding frequency $\omega $.
Such functions will correspond to the LSC  and other coherent flow features \citep[][]{Mezic13}. 

When dealing with systems with quasi-periodic or chaotic behavior exhibiting non-isolated eigenvalues and/or continuous spectrum, the eigenvalue 
problem for the raw generator $V$ is numerically ill-conditioned. Instead, one may seek approximate Koopman eigenvalues and eigenfunctions 
associated with a regularized generator  
%-------------------------------
\begin{equation}
\label{eqL}
L = V - \zeta \Delta, 
\end{equation}
%-------------------------------
where $ \zeta>0$ is a small regularization parameter, and $ \Delta $ a positive-semidefinite diffusion  operator  \citep[][]{GiannakisEtAl15,Giannakis17,DasGiannakis17}. Under relatively mild assumptions, $L $ has a point spectrum only \citep[][]{FrankeEtAl10}, and its eigenfunctions $ \psi $ approximate the eigenfunctions of $ V $ as $\zeta\to 0$. Moreover, the imaginary part of the eigenvalues $ \lambda $ of $L $ approximate the Koopman eigenfrequencies $ \omega $. The real part of $ \lambda $ is non-positive by construction and is related to a Dirichlet energy (roughness) of $ \psi $, as discussed below. 

\subsection{\label{secNLSA}Data-driven basis from nonlinear Laplacian spectral analysis}
In order to solve the eigenvalue problem for $ L $, we first specify an orthonormal basis of $ L^2( A, \mu ) $ which can be approximated from data sampled on the orbit $ \{ x_n \} $. Here, we build our data-driven basis by applying NLSA to the velocity-field snapshots $ \bm u_{-Q+1}, \ldots, \bm u_{N-1} $, given by $ \bm u_n = F( x_n ) $, where $F$ is the observation function (projection to velocity field data), defined in \eqref{proj}.  Applied to this dataset, NLSA approximates an ergodic, compact Markov operator $ P : L^2( A, \mu ) \mapsto L^2( A, \mu ) $ with real eigenvalues $ \Lambda_k $ and a corresponding set of orthonormal eigenfunctions $ \phi_k $. This operator is approximated by an $ N \times N $ Markov matrix $ \mathsf{ P } = [ P_{ij} ] $ with  $ \sum_{j=0}^{N-1} P_{ij} = 1 $, whose elements are computed through a symmetric kernel function 
$ K : X \times X \mapsto \mathbb{ R }_+ $ operating on delay sequences of velocity field snapshots. Given an arbitrary pair of states 
$ x, x' \in X $ with $ x = ( \bm{u}, T ) $ and $ x' = ( \bm{ u }', T' ) $, we define 
%-------------------------------
\begin{equation}
 K( x, x' ) =e^{-d^2_Q(x,x')/\epsilon} \quad\mbox{with}\quad 
 d^2_Q( x, x' ) =  \frac{ 1 }{ Q } \sum_{q=0}^{Q-1} \lVert F( \Phi_{-q\tau}( x ) ) - F( \Phi_{-q\tau}( x' ) ) \rVert^2_{L^2}\,,
\label{kernel}    
\end{equation}
%-------------------------------
where $ Q $ is a non-negative integer parameter (the number of delays), $ \epsilon $ a positive kernel bandwidth parameter, and $ \lVert \cdot \rVert_{L^2} $ denotes the norm in velocity field space $ L^2( \Omega; \mathbb{ R}^3 ) $. Moreover, $ d_Q $ is a distance function based on data in delay-coordinate space.  Intuitively, $K( x, x' )$ 
can be thought of as a pairwise measure of similarity between $ Q $-element sequences of velocity fields. In particular, for two states $ x_i $ and $ x_j $ 
in the training data, we have 
\begin{displaymath}
K_{ij} := K( x_i,x_j ) = \exp\left( - \frac{1}{\epsilon Q}\sum_{q=0}^{Q-1} \lVert \bm{ u }_{i-q} - \bm{ u }_{j-q} \rVert^2_{L^2} \right) \,.
\end{displaymath}
In NLSA, the matrix $ \mathsf{ P } $ is formed by performing the normalization procedure introduced in the diffusion maps algorithm to the pairwise kernel values, 
namely \citep[][]{CoifmanLafon06,BerrySauer16b}, 
\begin{equation}
    \label{eqPMat}
  P_{ij}=\frac{K_{ij}}{\left(\sum^{N-1}_{n=0} K_{in}q_n^{-1/2}\right)q_j^{1/2}} \quad\quad\mbox{with}\quad\quad q_i=\sum^{N-1}_{n=0} K_{in}\,.
\end{equation}

Let now $ f : X \mapsto \mathbb{ C } $ be a continuous function. Given the finite dataset $ \{ x_n \}_{n=0}^{N-1} $, we represent $f$ 
by the column vector $ \bm f = ( f( x_0 ), \ldots, f( x_{N-1} ) )^\top \in \mathbb{ C }^N $, whose components are equal to the values of $ f $ at the 
sampled states. It can be shown \citep[e.g.,][]{DasGiannakis17} that for every such function $ f$ the matrix-vector product  $ \bm g = \mathsf{ P } \bm f $ 
is an ergodic  average that approximates the action $ P f $. This, in conjunction with compactness of $ P$,  implies that the nonzero eigenvalues and corresponding eigenvectors, $ ( \Lambda_k, 
\bm \phi_k ) $, of $ \mathsf{ P } $ approximate those of $ P $ \citep[][]{VonLuxburgEtAl08}. Moreover, as $N \to \infty $, the set $  \{ \bm \phi_k \} $ 
converges in an appropriate sense to a basis for a closed subspace $ \tilde{\mathcal{ D }} $ of  $ L^2( A, \mu ) $. One further verifies that 
the eigenvalues of $ \mathsf{ P } $ are real, and admit the ordering $  1 = \Lambda_0 > \Lambda_1 \geq \Lambda_2 \geq \cdots $. We will use this ordering throughout the paper. In addition, the eigenvectors $\bm \phi_k $ are orthogonal with respect to the weighted inner product
\begin{equation}
    \langle \bm f, \bm g \rangle_{\mathsf{P}} = \sum_{i=0}^{N-1} f^*_i g_i w_i, \quad \bm f = ( f_0, \ldots, f_{N-1} )^\top, \quad \bm g = ( g_0, \ldots, g_{N-1} )^\top,  
    \label{eqInnerProd}
\end{equation}
where the weights $ w_i $ are equal to the components of the stationary distribution $\bm w = ( w_0, \ldots w_{N-1} )$ of $ \mathsf{P}$, satisfying $ \bm w \mathsf{P} = \bm w $ and $ \sum_{i=0}^{N-1} w_i = 1 $.

A key property of $ \mathsf{ P }$ constructed via the NLSA kernel in~\eqref{kernel} is that as the number of delays $ Q $ grows, $ \tilde{\mathcal{ D }} $ becomes 
invariant under the action of the Koopman operator $ U_t $, and is also contained in the so-called discrete subspace $ \mathcal{ D } $ of $ U_t $ 
spanned by all Koopman eigenfunctions \citep[][]{DasGiannakis17} . We therefore have the invariant inclusions $ \tilde{\mathcal{D}} \subseteq 
\mathcal{D} \subseteq L^2(A,\mu) $. In particular, every Koopman eigenfunction $ \psi \in \tilde{\mathcal{D}} $ can be approximated by a linear 
combination $ \bm \psi = \sum_{k=0}^{\ell-1} c_k \bm \phi_k $ of eigenvectors of $ \mathsf{ P } $ for some spectral truncation parameter 
$ \ell \leq N - 1 $, and in a limit $\ell,Q,N \to \infty $, $ \bm \psi $ converges to $\psi $. 

Another important property of $ \mathsf{ P } $ is that it induces a measure of roughness of observables through its eigenvalues. Specifically, 
to each $ \bm \phi_k $  we assign the quantity $ \eta_k $ with 
\begin{equation}
\eta_k= \frac{1}{\epsilon} \left(1 - \Lambda_k^{-1} \right)\,,
\label{etak}
\end{equation}
and $ \eta_0, \eta_1, \ldots $ forms an increasing sequence 
with $ \eta_0 = 0 $. Here, the parameter $\epsilon$ stands again for the kernel bandwidth (see~\eqref{kernel}). If the invariant set $ A $ has a Riemannian manifold structure, the $ \eta_k $ converge in a suitable asymptotic limit $ \epsilon \to 0 $ and $ N \to \infty $, and up to a proportionality constant, to the values  $ E_k = \mathcal{ E }( \phi_k ) $ of the Dirichlet energy functional $ \mathcal{ E }( \phi_k ) = \int_A \lVert \grad \phi_k \rVert^2 \, d\mu $, where $ \phi_k $ is again the eigenfunction of $ P $ to which $ \bm \phi_k $ converges \citep[][]{CoifmanLafon06}.  In this limit, the $ E_k  $ form  together with the $ \phi_k$ eigenvalue-eigenfunction pairs of the Laplacian $ \Delta = - \divr \grad $ on the manifold.  Given our convention of ordering the $\eta_k$ in increasing order, for every non-negative integer $\ell$, $ \{ \phi_0, \ldots, \phi_{\ell-1} \} $ corresponds to an $\ell$-element orthonormal set of functions in $ L^2(A,\mu) $ with the least Dirichlet energy.

Intuitively, the Dirichlet energy can be thought of as a measure of ``roughness'', or spatial variability, of functions on a manifold $A$. In particular, a highly oscillatory function $ f : A \mapsto \mathbb{R} $ has, on average, large values of its gradient norm $\lVert \grad f \rVert$, and thus large Dirichlet energy $\mathcal{E}(f)$, whereas $\mathcal{E}(f)$ will be small if $ f $ is spatially smooth. Note that the Dirichlet energy should not be understood as a kinetic or thermal energy. Rather, if one were to think of function $ f $ as describing the displacement of a linearly elastic membrane on $ A$, $\mathcal{E}(f)$ would measure the elastic potential energy stored in the membrane.  

While the invariant set $ A $ of our turbulent convection flow is not a smooth manifold, and we only have access to finitely many snapshots, one can still think of the $\eta_k$ from~\eqref{etak} as Dirichlet energies, measuring the roughness of the corresponding eigenvectors $ \bm\phi_k $.  In particular, observe that because $ \mathsf{ P } $ is Markov, $ \bm \phi_0 = ( 1, 1, \ldots, 1 )^\top \in \mathbb{ C }^N  $ is the constant eigenvector which gets assigned zero Dirichlet energy, and $ \bm \phi_1, \bm \phi_2,\ldots $ are mutually orthogonal eigenvectors with respect to the inner product in~\eqref{eqInnerProd} of increasingly higher energy. Moreover, despite $A$ lacking a manifold structure, it is still possible to construct a self-adjoint, positive-semidefinite, unbounded operator $  \Delta : D(\Delta) \mapsto L^2(A,\mu)  $, analogous to the Laplacian on Riemannian manifolds. This operator has a purely discrete spectrum, and acts on functions $ f = \sum_{k=0}^\infty c_k \phi_k $  through 
the formula 
%-------------------------------
\begin{equation}
\label{eqDelta}
\Delta f =\sum_{k=0}^\infty E_k c_k \phi_k,
\end{equation}
%-------------------------------
where the $ E_k $ are quantities to which the $ \eta_k $ converge as $ N \to \infty$ \citep[][]{DasGiannakis17}. The data-driven analog $ \mathsf{D} $ of 
this operator is then defined through 
%-------------------------------
\begin{equation}
\mathsf{D} \bm f = \sum_{k=0}^{N-1}\eta_k c_k \bm \phi_k \,, 
\end{equation}
%-------------------------------
where $\bm f = \sum_{k=0}^{N-1} c_k \bm\phi_k$. 

From a numerical standpoint, it should be noted that the $ ( \Lambda_k, \bm \phi_k ) $ can be computed stably and efficiently by solving the 
eigenvalue problem for the $ N \times N $ Markov matrix $ \mathsf P $. In particular, at large sample numbers $ N $, that  matrix can be made sparse 
by zeroing all but the largest $ k_\text{nn} \ll N $ elements in each row of the un-normalized kernel matrix $ \mathsf{K} = [K_{ij} ] $, and symmetrizing 
the resulting sparse matrix. Moreover, the computation to form $ \mathsf{ P} $ scales linearly with the data space dimension $ d $ (in this case, the number 
of velocity field gridpoint values), and is trivially parallelizable. Indeed, once $ \mathsf{K} $ has been computed, the cost of forming $ \mathsf{ P } $ and 
computing $  ( \Lambda_k, \bm\phi_k) $, as well as the cost of the Koopman eigenvalue problem described in section~\ref{secKoopEig} ahead, is 
independent of the ambient space dimension. In the problem studied here, $ N = \text{10,000} $ is small-enough so as not to require sparsification of 
$ \mathsf{ P } $, but this would become important at larger sample numbers (e.g., for the analysis of convection at higher Rayleigh numbers). However, 
the number of degrees of freedom of the velocity field in the DNS model has the value $ d = 3 \times 128^3 \approx 6.3 \times 10^6 $, which is more than 
600 times larger than $ N $, 
and the favorable computational cost of our approach with respect to $ d $ becomes crucial.

\subsection{\label{secKoopEig}Galerkin method for the Koopman eigenvalue problem}
    
We compute approximate Koopman eigenfunctions through a Petrov-Galerkin method for the eigenvalue problem of the regularized generator $L$
in~\eqref{eqL}, defined using the diffusion operator $\Delta $ in~\eqref{eqDelta}.  This scheme will be formulated in a data-driven basis consisting 
of eigenfunctions of $ \mathsf{ P } $. First, for any $ p \in \mathbb{ N } $, we introduce the Sobolev spaces   
%-------------------------------
\begin{displaymath}
H^p = \left \{ f = \sum_{k=1}^\infty c_k \phi_k \in L^2(A,\mu) : \sum_{k=1}^\infty E_k^p \lvert c_k^2 \rvert < \infty \right \},
\end{displaymath}
%-------------------------------
and equip these spaces with the inner products 
%-------------------------------
\begin{displaymath}
\langle f, g \rangle_{H^p} = \sum_{k=1}^\infty ( 1 + E_k + \ldots + E_k^p ) c_k^* d_k\,, 
\end{displaymath}
%-------------------------------
and thus the norms $\lVert f \rVert_{H^p} = \langle f, f \rangle^{1/2}_{H^p}$, where 
%-------------------------------
\begin{displaymath}
f =\sum_{k=1}^\infty c_k \phi_k \quad\mbox{and}\quad g = \sum_{k=1}^\infty d_k \phi_k \,.
\end{displaymath} 
%-------------------------------
Note that the infinite-dimensional spaces $H^p$ contain functions of higher regularity than $L^2 $ by imposing a condition on the rate of decay of the expansion coefficients $ c_k $, which becomes stronger as $p $ grows. Moreover, all $ H^p $ spaces contain zero-mean functions since they are $ L^2 $-orthogonal to the constant function $ \phi_0 $. In what follows, we will only require the spaces with $ p \leq 2 $.  

With these definitions in place, the eigenvalue problem for  $L$ is approximated by the following well-posed 
regularized Koopman eigenvalue problem in weak form:
%-------------------------------
\begin{equation}
{\cal L}(f,\psi):=\langle f,V\psi\rangle_{H^0}-\zeta\langle f,\Delta\psi\rangle_{H^0}=\lambda\langle f,\psi\rangle_{H^0}, \quad \forall f \in H^0.
\label{def1}
\end{equation}
%-------------------------------
Here, $ \lambda \in \mathbb{ C } $ and $ \psi \in H^2 $ are weak eigenvalues and eigenfunctions of $ L $. Furthermore, $f$ is a test function, and ${\cal L}:H^0\times H^2\mapsto \mathbb{C}$ is bounded. 

Passing to a data-driven approximation of the eigenvalue 
problem involves two steps, namely (i) approximation of the trial and test spaces by appropriate ``data-driven'' Sobolev spaces, 
and (ii) approximation of the action of the generator $ V $ on functions by finite differences or matrix logarithms. 

Our data-driven Sobolev spaces are defined in direct analogy with their infinite-dimensional counterparts as $ H_N^p := \spn 
\{ \bm\phi_1, \ldots, \bm \phi_{N-1} \} $, and are equipped with the inner products and norms 
\begin{displaymath}
\langle \bm f, \bm g \rangle_{H^p_N} = \sum_{k=1}^{N-1} ( 1 + \eta_k + \ldots + \eta_k^p ) c_k^* d_k, \quad \lVert \bm f \rVert_{H^p_N} = \langle \bm f, \bm f \rangle^{1/2}_{H^p_N},
\end{displaymath}
respectively, where 
\begin{displaymath}
\bm f = \sum_{k=1}^{N-1} c_k \bm \phi_k \quad\mbox{and}\quad \bm g = \sum_{k=1}^{N-1} d_k \bm \phi_k\,.
\end{displaymath}
Note that $\dim(H^p_N)=N - 1 $, and that ${\bm \phi}_0$ (which is not included in our definition of $ H^p_N$) has constant elements. Moreover, the inner product of $H^0_N$ is equivalent to the weighted inner product in~\eqref{eqInnerProd}.
We equip the data-driven Sobolev spaces $H_{N}^p$ with the normalized basis functions $ \bm \phi_k^{(p)} =\bm \phi_k / \eta_k^{p/2} $ 
for $ k \in \{ 1, 2, \ldots, N -1 \}$.  The normalization factors $\eta_k$ are given by \eqref{etak}.
These basis functions have the important property that for any $ \bm f = \sum_{k=1}^{N-1} c_k \bm \phi_k^{(p)} $, converging as $N \to \infty $ to $ f \in H^p $, the $ H^p_N $ norm can be  bounded by $ C \lVert \bm c \rVert $, where $ C $  is a constant independent of $ N $, and $ \lVert \bm c \rVert $ the 2-norm of the vector of 
expansion coefficients $ \bm c = ( c_1, \ldots, c_{N-1} ) \in \mathbb{ C }^{N-1} $. This means that as $ N \to \infty $, $  \{ \bm \phi^{(p)}_k \}_{k=1}^{N-1} $  converges to a basis of $ H^p $, whereas the unnormalized basis functions  $ \bm \phi_k $ do not.   

To approximate the action of the generator on functions, recall from section~\ref{secDefKoop} that $ Vf $ corresponds to a ``time derivative'' of 
observable $f$ along the dynamical flow, and that $V$ is related to the Koopman operator $U^t$ via exponentiation, $U^t = \exp(tV)$. For a sufficiently smooth function $ f : X \mapsto \mathbb{ C }  $, these facts respectively suggest that we can approximate $ V f $ through finite differences in time or via matrix logarithms. As a concrete example of the former approach, the quantity 
\begin{displaymath}
  V_\tau f = \frac{ U_{\tau} f - U_{-\tau  } f }{ 2 \tau } 
\end{displaymath}    
is a second-order, central finite-difference approximation of $V f$ for the sampling interval $ \tau $. Evaluating this expression at a state $ x_n $ of the training data, we have
\begin{displaymath}
  V_\tau f( x_n ) = \frac{ f( x_{n+1} ) - f( x_{n-1} ) }{ 2\tau}.
\end{displaymath}
Moreover, since the components $ \phi_{nk} $ of the basis vectors $ \bm \phi_k = ( \phi_{0k}, \ldots, \phi_{N-1,k} ) $ correspond to the values of a function at $ x_n \in X $ (see section~\ref{secNLSA}), we can define an approximate generator $ \mathsf V_\tau : H^0_N \mapsto H^0_N $ based on the action of $ V_\tau $ on the $ \bm \phi_k $ basis elements. Specifically, introducing the vectors $ \bm \phi_k' = ( \phi_{0k}', \ldots, \phi_{N-1,k}' ) $ with elements
\begin{displaymath}
    \phi'_{nk} = \frac{ \phi_{n+1,k} - \phi_{n-1,k}}{ 2 \tau}, \quad 1 \leq n \leq N - 2,
\end{displaymath}
and $ \phi'_{0k} = \phi'_{N-1,k} =0 $, the action  $ \mathsf V_\tau \bm f = \bm g $ on an observable $ \bm f = \sum_{k=1}^{N-1} c_k \bm \phi_k \in H^0_N $ is given by $ \bm g = \sum_{j=1}^{N-1} d_j \bm \phi_j $, where
\begin{equation}
    \label{eqVFD}
    d_j = \sum_{k=1}^{N-1} V_{\tau,jk} c_k, \quad V_{\tau,jk} = \langle \bm \phi_j, \mathsf V_\tau \bm \phi_k\rangle_{H^0_N} =  \langle \bm \phi_j, \bm \phi_k' \rangle_{H^0_N}.  
\end{equation}
In the above, the quantities $ V_{\tau,jk} $ are the matrix elements of $ \mathsf{V}_\tau $ in the $ \{ \bm \phi_k \}_{k=1}^{N-1} $ basis of $ H^0_N $. Alternatively, to construct an approximation of $ V $ based on matrix logarithms, we proceed as above, replacing $ \mathsf V_\tau $ in~\eqref{eqVFD} with 
\begin{equation}
    \label{eqVLog}
    \mathsf{ V }_\tau = \tau^{-1} \log \mathsf U^\tau, \quad \mathsf{ U }^\tau = [ U^\tau_{jk} ], \quad U^\tau_{jk} = \langle \bm \phi_j, \bm \phi_k^{+} \rangle_{H^0_N},  
\end{equation}
where $ \bm \phi_k^+ = ( \phi_{0k}^+, \ldots, \phi_{N-1,k}^+ ) $ is the vector with elements
\begin{displaymath}
    \phi_{nk}^+ = \phi_{n+1,k}, \quad 1 \leq n \leq  N - 1,
\end{displaymath}
and $ \phi_{0,k}^+ = 0 $.

%---------------------------------------------------------------------
\begin{table}
\begin{center}
\begin{tabular}{lccc}
  & Symbol  &  Definition  &   Eigenvalue Problem \\
\hline
Koopman operator  &  $U_t$ &  $ U_t f = f \circ \Phi_t $& $U_t \psi=\mbox{e}^{i\omega t} \psi$  \\     \\                 
Koopman generator & $V$ & $ V f = \frac{ d U_t f }{ d t } \Big|_{t=0} $ & $V\psi=i\omega \psi$ \\ \\
Regularized Koopman & $L$ & $Lf=Vf-\zeta\Delta f$ & $L\psi=\lambda \psi$ \\
generator & & & \\
Sesquilinear form & ${\cal L}$ & ${\cal L}(f,\psi)=\langle f,V\psi\rangle_{H^0}-\zeta\langle f,\Delta\psi\rangle_{H^0}$ & ${\cal L}(f,\psi)=\lambda\langle f,\psi\rangle_{H^0}$ 
\\
associated with $L$ & & & \\
Data-driven, Galerkin- & $\hat{\cal L}$ & $\hat{\cal L}(\bm f,\bm \psi)=\langle \bm f,\mathsf V\bm \psi\rangle_{H^0_{N}}-\zeta\langle \bm f,\mathsf D\bm\psi\rangle_{H^0_{N}}$ 
& $\hat{\cal L}(\bm f,\bm \psi)=\lambda\langle \bm f,\bm \psi\rangle_{H^0_{N}}$ \\
approximated $ \mathcal{L} $ & & & 
\end{tabular}  
\caption{Summary of the several levels of description towards the data-driven and Galerkin-approximated 
eigenvalue problem for the regularized Koopman generator in weak form. The step from $\mathcal{L}$ to $\hat{\cal L}$ requires 
the generation of a data-driven basis which consists of the eigenfunctions of the Markov operator $\mathsf P$ obtained by NLSA.}
\label{Tab1}
\end{center}
\end{table}
%---------------------------------------------------------------------------

With these definitions, we  pose the following variational eigenvalue problem, which is a data-driven analog of the regularized Koopman 
eigenvalue problem in \eqref{def1}. Fixing a spectral truncation parameter $ \ell \leq N - 1  $, the problem consists of finding eigenvalues 
$ \lambda \in \mathbb{ C } $ and eigenfunctions $ \bm \psi \in \spn \{ \bm \phi_1^{(2)}, \ldots, \bm \phi_\ell^{(2)} \} $  such that for all 
$ \bm f \in \spn\{ \bm \phi_1, \ldots, \bm \phi_\ell \}  $, 
\begin{equation}
   \hat {\cal L}( \bm f, \bm \psi ) = \lambda \langle \bm f, \bm \psi \rangle_{H^0_N}
\label{def2}   
\end{equation}
holds, where $ \hat {\cal L} : H_N^0 \times H_N^2 \mapsto \mathbb{ C } $ is the sesquilinear form defined as
\begin{displaymath}
    \hat {\cal L}( \bm f, \bm \psi ) = \langle \bm f, \mathsf V_\tau \bm \psi \rangle_{H^0_N} - \zeta \langle \bm f,  \mathsf D \bm \psi \rangle_{H^0_N},  
\end{displaymath}
and $\mathsf V_\tau$ an approximation of the generator, obtained either via the finite difference, or matrix logarithm approaches in~\eqref{eqVFD} and~\eqref{eqVLog}, respectively. Numerically, this variational problem is equivalent to solving a matrix generalized eigenvalue problem
\begin{equation}
    \mathsf L \bm c_k = \lambda_k \mathsf B \bm c_k, \quad \lambda_k \in \mathbb{C},
  \label{evp}
\end{equation}
where $ \mathsf L $ and $ \mathsf B $ are $ \ell \times \ell $ matrices with elements
\begin{equation}
    \label{eqLBMat}
    L_{ij} = \hat {\cal L}( \bm \phi_i, \bm \phi_j^{(2)} ) = V_{\tau,ij} \eta_j^{-1} - \zeta \delta_{ij},  
  \quad B_{ij} = \langle \bm\phi_i, \bm \phi_j^{(2)} \rangle_{H^0_N} = \eta_i^{-1} \delta_{ij},
\end{equation} 
respectively, and $ \bm c_k = ( c_{1k}, \ldots, c_{\ell k} )^\top $ is a column vector in $ \mathbb{ C}^\ell $ containing the expansion coefficients of the solution $ \bm \psi_k $ in the $ \{ \bm \phi_j^{(2)} \} $ basis of $ H^2_N $, viz. $ \bm \psi_k = \sum_{j=1}^{\ell} c_{jk} \bm \phi_j^{(2)} $. Note that matrix eigenvalue problems of this class are encountered frequently in finite element methods for elliptic partial differential equations \citep{BabuskaOsborn91}, where the matrices $\mathsf{L}$ and $\mathsf{B}$ are referred to as stiffness and mass matrices, respectively. 

It can be shown that under mild assumptions on the convergence of the $ \eta_k $ to the Dirichlet energies $ E_k $, for any $ \ell $, 
the solutions of the eigenvalue problem in (\ref{def2}) converge as $ N \to \infty $ to the solutions of the problem in (\ref{def1}), 
restricted to the $ \ell $-dimensional subspace of $ H^2 $ spanned by $ \{ \phi_1, \ldots,\phi_{\ell} \} $. Note again that it is important 
to work with the normalized basis $ \{ \bm \phi_k^{(2)} \} $ basis of $ H^2_N $ as opposed to the un-normalized basis  $ \{ \bm \phi_k \} $ 
in order to ensure good conditioning of our scheme at large $ \ell $. This is reflected from the fact that $ \mathsf D $ is represented in this 
basis by the identity matrix, whose condition number is 1 for all $ \ell $. In practice, one typically works with $ \ell \ll N - 1 $, so 
that~\eqref{evp} can be feasibly solved at large sample numbers. As with the NLSA basis functions from section~\ref{secNLSA}, 
we order our solutions $ ( \lambda_k, \bm \psi_k) $, $ k \in \{ 1, \ldots, \ell \} $, in order of increasing Dirichlet energy, 
\begin{equation}
    \label{eqDirichlet}
    \hat{\mathcal{E}}(\bm\psi_k) := \frac{ \langle \bm \psi_k, \mathsf D \bm \psi_k \rangle_{H^0_N}  }{ \lVert \bm \psi_k\rVert^2_{H^0_N} } = \frac{\sum_{i=1}^\ell \lvert c_{ik} \rvert^2 / \eta_i }{ \sum_{j=1}^\ell \lvert c_{jk} \rvert^2 / \eta_j^2 }.
\end{equation} 
As stated in section~\ref{secDefKoop}, the Koopman eigenfunctions corresponding to nonzero eigenvalues form complex conjugate pairs (with equal Dirichlet energies), and this is usually the case for the data-driven $ ( \lambda_k, \bm \psi_k ) $.  However, due to the presence of diffusion, it is possible that $ \mathsf L $ has purely real, negative eigenvalues with corresponding real eigenfunctions. We will, in fact, encounter such a solution in section~\ref{secResults} ahead.  Note that the Koopman eigenvalue problem also has the trivial (constant) solution $ \lambda_0 = 0 $ and $ \bm \psi_0 = \bm \phi_0 $, with vanishing Dirichlet energy, but our numerical eigenvalue problem does not yield this solution since our trial space is orthogonal to $ \bm \psi_0 $.  

Table \ref{Tab1} summarizes the different levels of description starting from the original Koopman operator eigenvalue problem 
to data-driven and Galerkin approximated eigenvalue problem for the regularized Koopman generator. 

\subsection{\label{secRecon}Spatiotemporal reconstruction}

Solving the eigenvalue problem in~\eqref{evp} results in the data-driven complex Koopman eigenvalues $\lambda_k \in \mathbb{C}$ and the corresponding eigenfunctions $ \bm \psi_k \in \mathbb{C}^N$, with $1 \leq k \leq \ell \ll N -1 $ and $N$ being the number of turbulent convection snapshots. 
The spatiotemporal patterns corresponding to the eigenfunctions are computed
through the standard approach employed in singular spectrum analysis algorithms (the analog of POD in delay-coordinate space) by \cite{GhilEtAl02}, which is also employed in NLSA. Here, we describe the reconstruction procedure for the velocity field itself, but note that the same approach can be used to reconstruct other observables such as the temperature field or combined quantities of both fields. 

First, given a lead/lag $ t' \in \mathbb{ R }$, we compute the  projections $ \bm A_k( t') $ of the observable $ F $ onto the Koopman eigenfunctions $ \psi_k $, 
\begin{equation}
  \label{eqAHat0}
    \bm A_k( t' ) = \langle \psi_k, F \circ \Phi_{t'} \rangle_{H^0} = \int_A \psi^*_k( x ) F( \Phi_{t'}( x ) )  \, d\mu(x).
\end{equation}  
Note that for each time $ t' $, $  \bm A_k( t' ) $ is a velocity field snapshot in $ L^2( \Omega; \mathbb{ R }^3 ) $. In applications, we evaluate this quantity at the discrete times $ t'_q = q \tau $ with $ q \in \mathbb{ Z} $, and approximate the integral with respect to the invariant measure $ \mu $ in \eqref{eqAHat0} by time averages. This leads to a complex-valued vector field  given by
\begin{equation}
  \label{eqAHat}
  \hat{\bm A }_k( q \tau ) = \frac{ 1 }{ N' } \sum_{n={0}}^{N'-1} \psi^*_{nk}\bm u_{n+q} \quad\quad\mbox{with}\quad\quad N' = \min\{ N, N+q \}.  
\end{equation}
In~\eqref{eqAHat}, we made use of the fact that the observable $F$ evaluated on a state $(\bm u, T)$ of the Boussinesq system is simply equal to the velocity field $\bm u $; see section 2. In particular, ${\bm u}_{n+q}$ is 
the velocity field of the convection flow sampled at time $ t = ( n + q ) \tau $, and moreover $ \psi_{nk} $ the $ n  $-th component of the Koopman eigenvector $ \bm \psi_k = ( \psi_{0k}, \ldots, \psi_{N-1,k} )^\top $. In the Koopman operator literature, $ \hat{\bm A}_k( 0 ) $ is referred to as the Koopman mode associated with the observable $ F $ \citep{Mezic05, BudisicEtAl12}.

The corresponding data-derived sequence of (complex-valued) spatiotemporal velocity field patterns associated with the pair $ ( \bm \psi_k, \hat{ \bm A }_k ) $ is given by
\begin{equation}
  \label{eqPsiRec}
  \hat {\bm u}^{(k)}_n = \frac{ 1 }{Q' } \sum_{q=0}^{Q'-1} \hat{\bm A}_k( -q\tau ) \psi_{n+q,k} \quad\mbox{with}\quad Q' = \min \{ Q, N - 1 - n \}, \quad 0 \leq n \leq N-1.  
\end{equation} 
Since the data-driven Koopman eigenfunctions usually form complex-conjugate pairs,  $ ( \bm \psi_k, \bm \psi_{k'} ) $,  we compute 
the sum $ \hat{\bm u}_k + \hat{\bm u}_{k'} $ to obtain a real-valued physical velocity field. As is evident from~\eqref{eqPsiRec}, a reconstruction at time $n\tau$ is a superposition 
of $Q$ snapshots. This is a result of the time-lagged embedding and different to a standard POD analysis.  

\section{\label{secResults}Koopman eigenfunctions and large-scale flow in the cell}
%------------------------------------------------------------------------------------------
\begin{figure}
    \centering
    \includegraphics[width=0.9\linewidth]{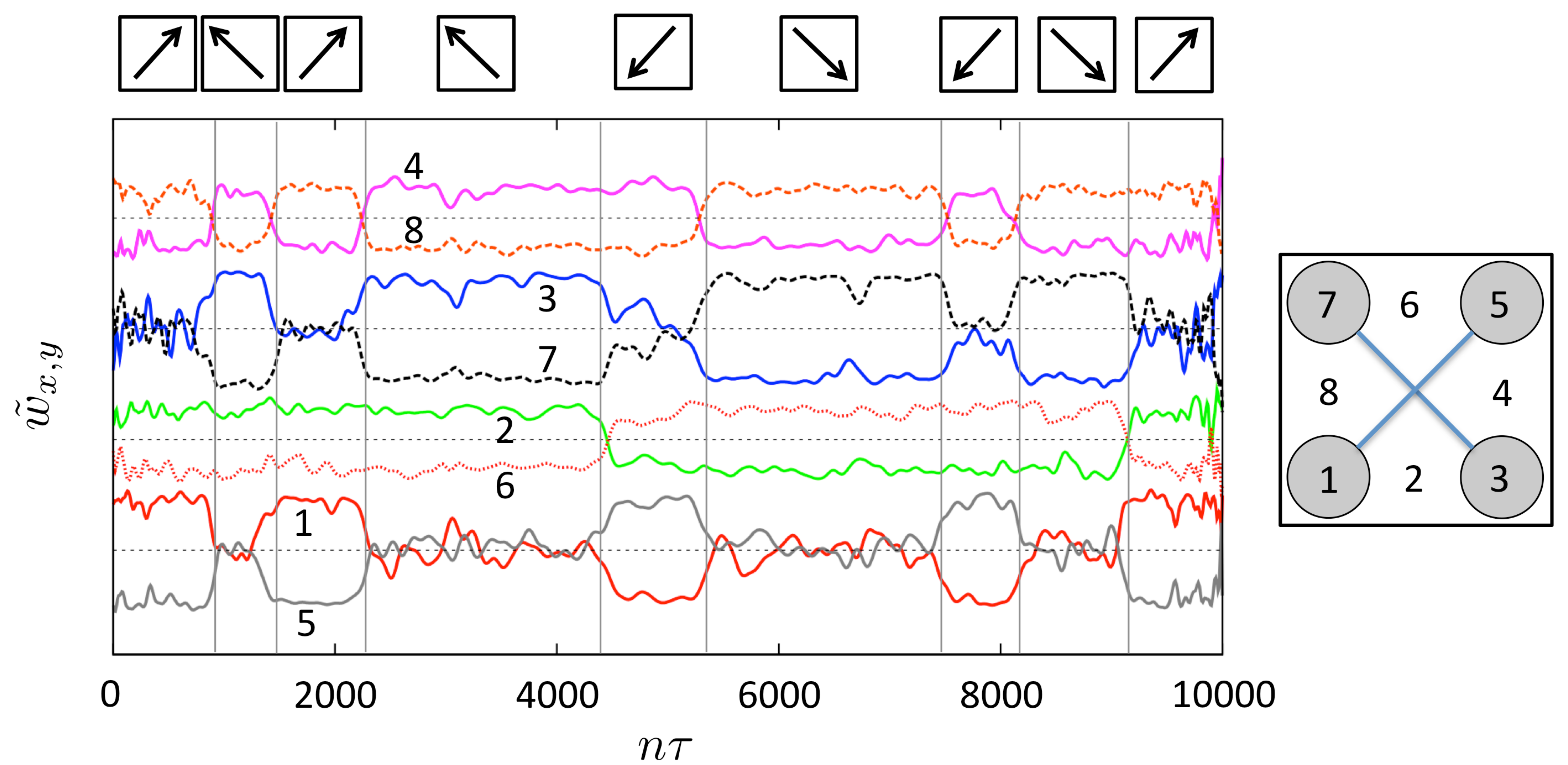}
    \caption{Time series of the mid-plane $(z=0)$, filtered vertical velocity $\tilde w_{x,y}$ from~\eqref{eqUZFilt}, sampled at four points near the corners of this 
cross section (labeled \#1, \#3, \#5, \#7) and at four points near the midpoints of the sides (labeled \#2, \#4, \#6, \#8) as shown in the sketch to the right of 
the plot. The time series is divided into several intervals (see vertical lines) which can be assigned to one the four diagonal large-scale flow states. These 
states are indicated by arrows in boxes (see top of figure) that stand for the direction of the large scale flow below the top plate of the cube. The time interval 
shown here is $ [0, 10000]$ in units of the free-fall time, which corresponds to $n\tau \in \{ 0,\dots, 10000 \} $. Within the panel, pairs of time series are shifted 
vertically to improve visibility. Dotted horizontal lines mark $\tilde{w}_{x,y} = 0 $ axes. The magnitude $|\tilde{w}_{x,y}| \le 0.5$.}
\label{Fig3}
\end{figure}
%------------------------------------------------------------------------------------------

In this section, we present the application of the Koopman eigenfunction analysis described in section~\ref{secKoopman} to the turbulent 
convection flow data from section~\ref{secBoussinesq}. The complex Koopman eigenfunctions, which we obtain from the numerical solution 
of the eigenvalue problem for the regularized generator $\mathsf L $ in~\eqref{evp}, can be separated into subgroups. As mentioned in 
subsection~\ref{secKoopEig}, ordering is done with respect to increasing Dirichlet energy;  the latter is found to be approximately the same 
for the  eigenfunctions within one group. Delay embedding (see~\eqref{kernel}) was performed using $ Q = 30 $ delays corresponding to a 
time interval of $30 \tau$. We selected the kernel bandwidth parameter $ \epsilon $ using the automatic tuning procedure described in 
\citet{BerryEtAl15} and \cite{Giannakis17}. The eigenvalue problem~\eqref{evp} was solved using a basis of  $ \ell = 2000 $ eigenfunctions 
$ \bm \phi_k $ from NLSA and the diffusion regularization parameter $ \zeta = 10^{-4} $. Thus, $\ell \ll N$ as discussed in section \ref{secKoopEig}. 

Table \ref{Tab2} summarizes the eigenfrequencies $\omega_k$ and Dirichlet energies $E_k$ of the leading few Koopman eigenfunctions, 
which we categorize as ``primary'' or ``secondary'' depending on their corresponding eigenfrequencies. In particular, primary eigenfunctions have 
low frequencies, $\omega_k = O(10^{-4})$, whereas secondary eigenfunctions have higher frequencies, $\omega_k \simeq 0.15$. In 
sections~\ref{secPrimary} to \ref{secSecondary}, we will discuss the large-scale properties of the convection flow represented by these two 
Koopman eigenfunction families. When convenient, we will employ the canonical polar coordinate $ \theta \in [ 0, 2\pi ) $ associated with the 
$(x,y)$ horizontal coordinates to characterize the orientation of LSC structures. That is, $ \theta $ is chosen such that the $(x,y) = (1,0.5)$ edge 
has $\theta = \pi/4$ orientation, and so on for the other sidewalls and edges. A sensitivity 
analysis of our results under changes of NLSA and Koopman operator parameters is included in section~\ref{secSensitivity}. 

%---------------------------------------------------------------------
\begin{table}
  \begin{tabular*}{\linewidth}{@{\extracolsep{\fill}}lll}
    Eigenfunction & Frequency $\omega_k$ & Dirichlet energy $E_k$\\
    \\
    Primary eigenfunctions \\
    $ \bm \psi_{1}, \bm \psi_{2} $ & $ \pm 2.97 \times 10^{-4} $ & 3.65 \\
    $ \bm \psi_3 $ & 0 & 5.80 \\
    \\
    Secondary eigenfunctions \\
    $ \bm \psi_{4}, \bm \psi_{5}$ & $ \pm $ 0.152 & 13.4 \\
    $ \bm \psi_{6}, \bm \psi_{7} $ & $ \pm $ 0.150 & 14.9 \\
    $ \bm \psi_{8}, \bm \psi_{9} $ & $ \pm $ 0.156 & 15.3 \\
    $ \bm \psi_{10}, \bm \psi_{11} $ & $ \pm $ 0.155 & 16.2 \\
  \end{tabular*}
   \caption{Properties of the first 11 data-driven Koopman eigenfunctions. Angular frequencies, $ \omega_k=\Im(\lambda_k) $, and Dirichlet energies, 
   $\hat{\mathcal{E}} (\bm \psi_k ) $, of the eigenfunctions $ \bm \psi_k $ are listed. The functions are sorted in order of increasing Dirichlet energy.
   The timescales associated with the frequencies $\omega_k$ are given by $T_k\simeq 2\pi/\omega_k$. Most of them are found in complex-conjugate pairs.}
  \label{Tab2}
\end{table} 
%---------------------------------------------------------------------------

\subsection{Time-averaging analysis at probe points}

Before coming to the Koopman analysis, we show in figure \ref{Fig3}  time series of the vertical velocity, measured at eight different 
simulation gridpoints in the midplane in their statistically steady state. This analysis is similar to the one conducted by \cite{Foroozani2017}. 
Four of these points are in the four corner regions, slightly away from the faces, and the other four points are located halfway between 
the corner points. Specifically, point \#1 in the $ \theta = 5 \pi / 4 $ corner is found  at $(x_1, y_1 ) =(0.078,-0.422)$,  point \#2 at $( 0.5,-0.422)$, and 
\#3 in the $ \theta = 7 \pi / 4 $  corner at $(0.922, -0.422)$.  The other five points $(x_4,y_4), \ldots, (x_8,y_8)$ follow correspondingly by symmetry 
in counterclockwise orientation. In addition, we applied a temporal and spatial smoothing of the strongly fluctuating signals to highlight 
the large-scale motion and to reduce the amount of available information. In detail, we substitute the vertical velocity  $u_z(x_i,y_i,0,t)=
w_{x,y}(t)$ at the monitoring points $ ( x_1, y_1 ), \ldots ( x_8, y_8 ) $ by 
%--------------------------------------------------------------------- 
\begin{equation}
  \label{eqUZFilt}
\tilde w_{x,y}(t)=\frac{1}{4}\left[\frac{w_{x+h,y}(t)+w_{x-h,y}(t)+w_{x,y+h}(t)+w_{x,y-h}(t)+4 w_{x,y}(t)}{2}\right]\,,
\end{equation}  
%--------------------------------------------------------------------- 
which corresponds to a local nearest-neighbor averaging on the simulation mesh with mesh width $h$. In addition, functions are smoothed with 
B\'{e}zier splines using a $ 10 T_f $ temporal window. The results confirm that all four diagonal states are visited over our analysis time series. In what follows, we will describe how Koopman eigenfunction analysis yields a representation of these states without requiring spatial or temporal smoothing, while also providing access to LSC patterns on shorter timescales (the secondary eigenfunctions) which would be difficult to extract using averaging approaches.

\subsection{\label{secPrimary}Primary eigenfunctions}

\begin{figure}
    \centering 
    \includegraphics[width=.75\linewidth]{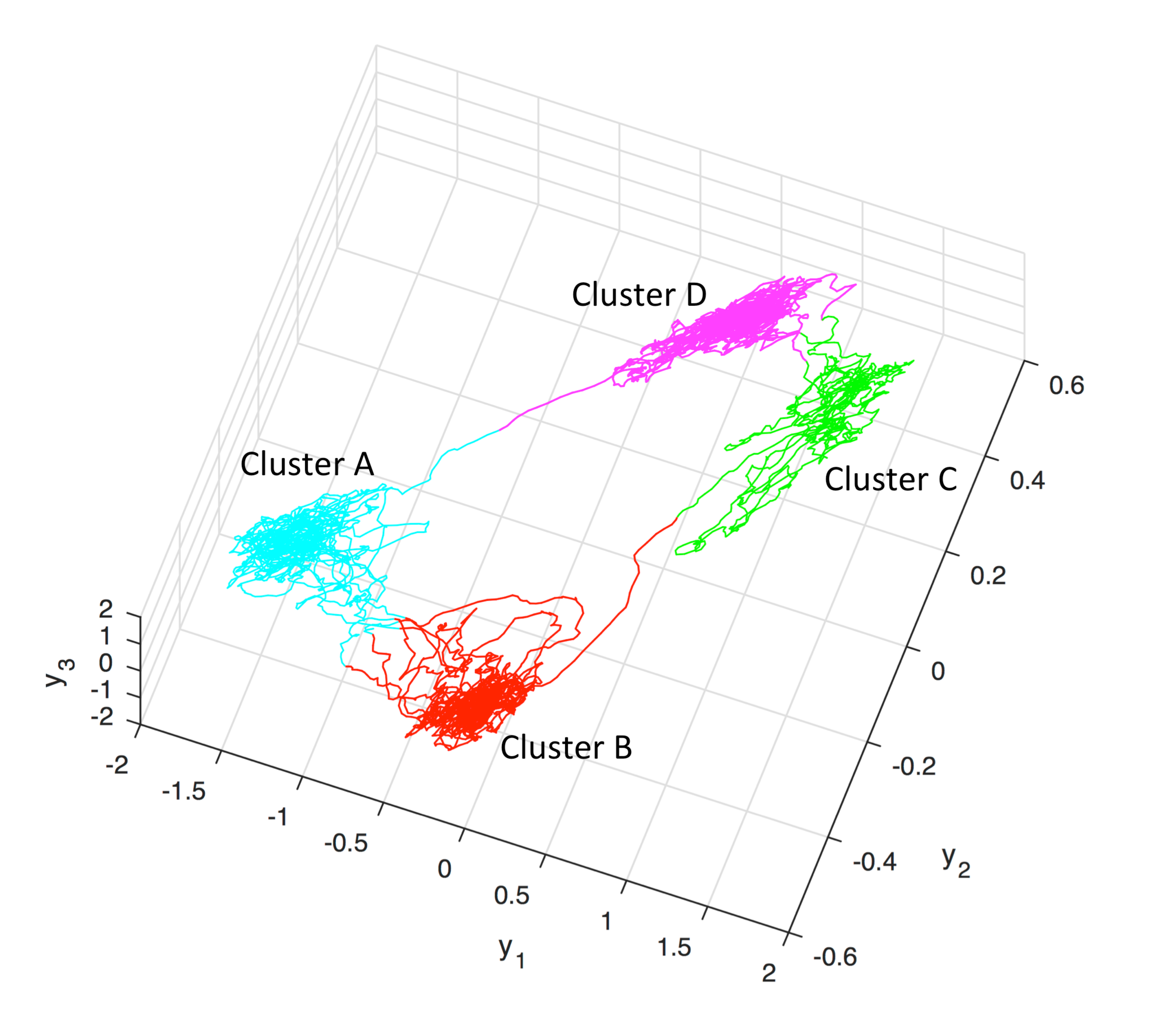}
\caption{Dynamical trajectory in the three-dimensional phase space with coordinates $(y_1(n),y_2(n),y_3(n))$, constructed from the primary Koopman eigenfunctions. The trajectory is sampled at the $ N = \text{10,000} $ analysis states at times $ t_n = n \tau $, where the sampling interval $ \tau $ is equal to one free-fall time unit. Four clusters can be assigned from a hierarchical $K$-means clustering performed on this point cloud. Notice the larger number of transitions between clusters A and B, or C and D, compared to B and C, or D and A.}     
\label{Fig4}
\end{figure}

The primary  eigenfunction family consists of the leading three data-driven Koopman eigenfunctions, $ \bm\psi_1, \bm \psi_2, \bm \psi_3 \in \mathbb{C}^N$, ordered with respect to increasing Dirichlet energy $\hat{\cal E}(\bm\psi_1)=\hat{\cal E}(\bm\psi_2) < \hat{\cal E}(\bm\psi_3)$.  
This family includes a complex-conjugate pair, $(\bm\psi_{1}, \bm\psi_{2})=\Re (\bm\psi_1)\pm i \Im (\bm\psi_1)$, 
and a purely real mode, $\bm\psi_3=\Re (\bm\psi_3)$. Thus, defining $ y_1( n ) = \Re(\psi_{n1})$, $ y_2( n )  = \Im(\psi_{n1})$, and 
$ y_3( n ) = \psi_{n3} $ with $n \in \{ 0, \ldots, N-1\}$, the evolution of the LSC configuration reconstructed from the primary eigenfunctions can be represented via a trajectory in a three-dimensional phase space with coordinates $ ( y_1, y_2, y_3 ) $,  displayed in figure~\ref{Fig4}. As stated in section~\ref{secDefKoop}, the trajectory is sampled at the $N=\text{10,000}$ 
temporally consecutive states $ x_n $ at times $ t_n =n \tau $, which are separated by one free-fall time unit, $ \tau = 1$,  from each other. Visually, it can 
readily be seen that the trajectory yields four clustered regions. These regions can be also identified quantitatively via $K$-means clustering performed on the $(y_1,y_2,y_3)$ coordinates. Residence times in these clusters can be well in excess of 1000 free-fall time units.
%------------------------------------------------------------------------------------------
%-------------------------------

\begin{figure}
    \includegraphics[width=\linewidth]{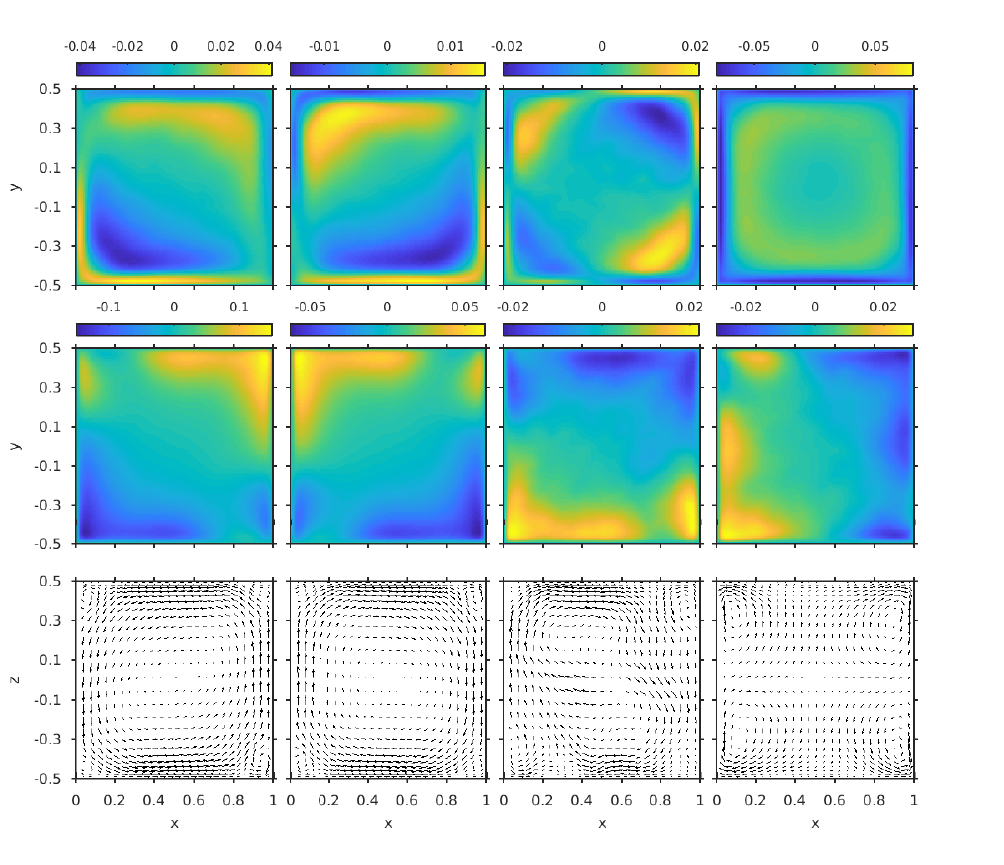}
    \caption{\label{Fig5}Projected velocity field snapshots $\hat{\bm A}_k(q \tau)$ at lag $ q= 0 $ for the primary Koopman eigenfunctions and the trivial (constant) Koopman eigenfunction, corresponding to a time average. The color plots in the first and second rows show the vertical velocity field components at horizontal cross-sections near the top ($z=0.45$) and at the middle ($z=0$) of the domain. The vector plots in the third row show the $ (\bm e_x, \bm e_z) $ velocity field components at a vertical cross-section perpendicular to $\bm e_y$ through the middle ($y=0$) of the domain. From left to right: $\Re \hat{\bm A}_1$, 
$\Im \hat{\bm A}_1$, $\hat{\bm A}_3$, and $\hat{\bm A}_0$.}
\end{figure}

In order to gain insight on the properties of the LSC states captured by the primary eigenfunctions, it is useful to examine their corresponding projected velocity field patterns $\hat{\bm A}_k(q\tau)$ from~\eqref{eqAHat}. Figure~\ref{Fig5} displays these patterns for $q=0$ for horizontal cross-sections near the top plate ($z=0.45$) and at the midplane ($z=0$), as well as for a vertical cross-section at $y=0$. We also show the time-averaged state, which corresponds to the $\hat{\bm A}_0(0)$ projection pattern associated with the trivial (constant) Koopman eigenfunction $\bm \psi_0 = ( 1, \ldots, 1 )^\top \in \mathbb{C}^N$. It is evident from the results in figure~\ref{Fig5} that the real and imaginary parts of $ \hat{ \bm A }_1(0) $ capture diagonal LSC patterns, with the circulation locked along one of the two diagonals of the square cross-section of the flow domain. On the other hand, at least when examined on the $z=0$ plane, the LSC patterns in  $\hat{ \bm A }_3(0)$ and the time-averaged state appear oriented along one of the two sidewall pairs. By examining $\hat{\bm A}_k(q\tau)$ at different values of $q$ (not shown here), it can be verified that the patterns in figure~\ref{Fig5} remain largely constant over the $Q=30$ delay embedding window. Due to this, in the ensuing discussion we will drop the $q\tau$ argument from $\hat{\bm A}_k(q\tau)$ in the interest of a simple notation.
%-------------------------------
%-------------------------------

Upon initial inspection, the vertical velocity field configuration in $ \Re \hat{\bm A}_1$ and $ \Im \hat{\bm A}_1$ suggests that these patterns are related by $\theta = \pi/2 $ (anticlockwise) rotations. From a dynamical symmetry standpoint, this is consistent with the fact that the Boussinesq system in a cubical domain is equivariant under the  $ \mathbb{Z}_4 $ cyclic group associated with rotations by integer multiples of $ \theta = \pi / 2 $. That is, if $( \bm u(\bm x, t), T(\bm x, t))$ is a solution, then $(\mathsf R_g \bm u( \mathsf R^{-1}_g \bm x, t), T( \mathsf R^{-1}_g \bm x, t))$ is also a solution, where $\mathsf R_g$ is a $3 \times 3$ rotation matrix representing group element $g \in \mathbb{Z}_4$. One can verify that if the system samples such solutions with equal probability in the course of dynamical evolution (more specifically, if the symmetry group action preserves the ergodic invariant measure of the dynamics), then the real and imaginary parts of the Koopman projected patterns $ \hat{\bm A}_k $ should span two-dimensional unitary representation spaces of the symmetry group, which implies in turn that $\Re \hat{ \bm A }_k$ and $\Im \hat {\bm A}_k$ should be relatable by symmetry operations \cite[for more details see, e.g.,][]{GiannakisEtAl17b}. The fact that $ \Re\hat{\bm A}_1$ and $\Im \hat{\bm A}_1$ are at least qualitatively relatable by a $\theta = \pi/2$ rotation is consistent with this picture. What is unexpected, however, is a difference in the velocity field magnitudes exhibited by $\Re\hat {\bm A}_1$ and $\Im \hat {\bm A}_1$. In particular, as can be seen in figure~\ref{Fig5}, typical vertical velocity field magnitudes are about a factor of two smaller in $\Im \hat{\bm A}_1$ than $\Re \hat{\bm A}_1$. This difference can be attributed to the fact that $ \lvert \Im \bm \psi_1 \rvert$ is typically about a factor of two smaller than $\lvert \Re \bm \psi_1 \rvert$, but, aside from our analysis not having sampled all LSC states in equal proportion, there is no obvious mechanism underlying such an asymmetry. We will return to the repercussions of this asymmetry to the properties of the reconstructed spatiotemporal patterns associated with the primary Koopman eigenfunctions below.

Next, turning to $ \hat {\bm A}_3$ and $\hat {\bm A}_0$ patterns, we first note that despite the $\mathbb{Z}_4$ equivariance of the Boussinesq system,  the time-averaged velocity field is not fully invariant under this symmetry group. This is manifested by the vertical velocity field pattern in $ \hat{\bm A}_0$ at the $z=0$ horizontal cross-section. This pattern exhibits a clear bipolar structure oriented in the $\theta = 0$ direction, indicating that the system does not visit all  $\mathbb Z_4$ symmetry states with equal probability in our analysis interval. Intriguingly, at this cross-section, the $\hat {\bm A}_3$ and $\hat {\bm A}_0$ patterns can be related via a $\theta = \pi/ 2$ rotation. However, elsewhere in the domain these patterns have pronounced differences, which cannot be accounted for by the $ \mathbb{Z}_4$ symmetry group. For instance, in the $z=0.45$ horizontal cross-sections in figure~\ref{Fig5}, the vertical velocity field associated with $\hat{\bm A}_3$ exhibits a quadrupolar pattern, whereas that associated with $\hat {\bm A}_0$ is unipolar.

We now examine the spatiotemporal reconstructions associated with the primary Koopman eigenfunctions. Due to the non-trivial structure of the time-averaged state with respect to $\mathbb{Z}_4$ rotations, and its relationship with $\hat {\bm A}_3 $ described above, in what follows we will include the time-averaged pattern in our reconstructions. That is, we consider the reconstructed velocity field
    \begin{equation}
        \label{eqPrimaryRec}
        \hat{\bm u}^\text{primary}_{n} := \hat{\bm u}^{(0)}_{n} + \hat{\bm u}^{(1)}_{n} + \hat{\bm u}^{(2)}_{n} + \hat{\bm u}^{(3)}_{n}, \quad 0 \leq n \leq N -1,
    \end{equation}
with $\hat{\bm u}^{(1)}_{n}$, $ \hat{\bm u}^{(2)}_{n}$, and $ \hat{\bm u}^{(3)}_n$ given by~\eqref{eqPsiRec}, and $ \hat{\bm u}_n^{(0)} = \hat {\bm A}_0( 0 ) $.
Note that the resulting velocity field projection is real-valued since Koopman eigenfunctions $\bm\psi_0$ and $\bm\psi_3$ are real-valued, and 
$\bm\psi_1$ and $\bm\psi_2$ form a conjugate-complex pair. We found that including $ \hat{\bm u}_0^{(n)}$ aids the physical interpretation of our results, as, without this pattern, our reconstructions would describe anomalies relative to the time-averaged state, which are somewhat cumbersome to visually interpret against a non-trivial time-averaged background. With this convention, owing to the fact that the $\hat{\bm A}_k(q\tau)$ patterns are nearly constant over the delay-embedding window for the primary Koopman eigenfunctions, our reconstructed velocity field patterns are, to a good approximation, given by the time-averaged state $ \hat{ \bm A }_0(0) $, plus linear combinations of the $ \Re \hat{\bm A}_1(0) $, $ \Im \hat{\bm A}_1(0) $, and $ \hat{\bm A}_3(0) $ patterns, with coefficients determined from the time averages of the $ y_1 $, $ y_2 $, and $y_3$ coordinates, respectively, over the time interval $ [ n\tau, n\tau + ( Q -1 ) \tau ] $. The nature of the resulting spatiotemporal patterns will therefore depend on the temporal relationships of the $y_k$, and particularly their clustering behavior, as we now discuss. 

\begin{figure}
    \includegraphics[width=\linewidth]{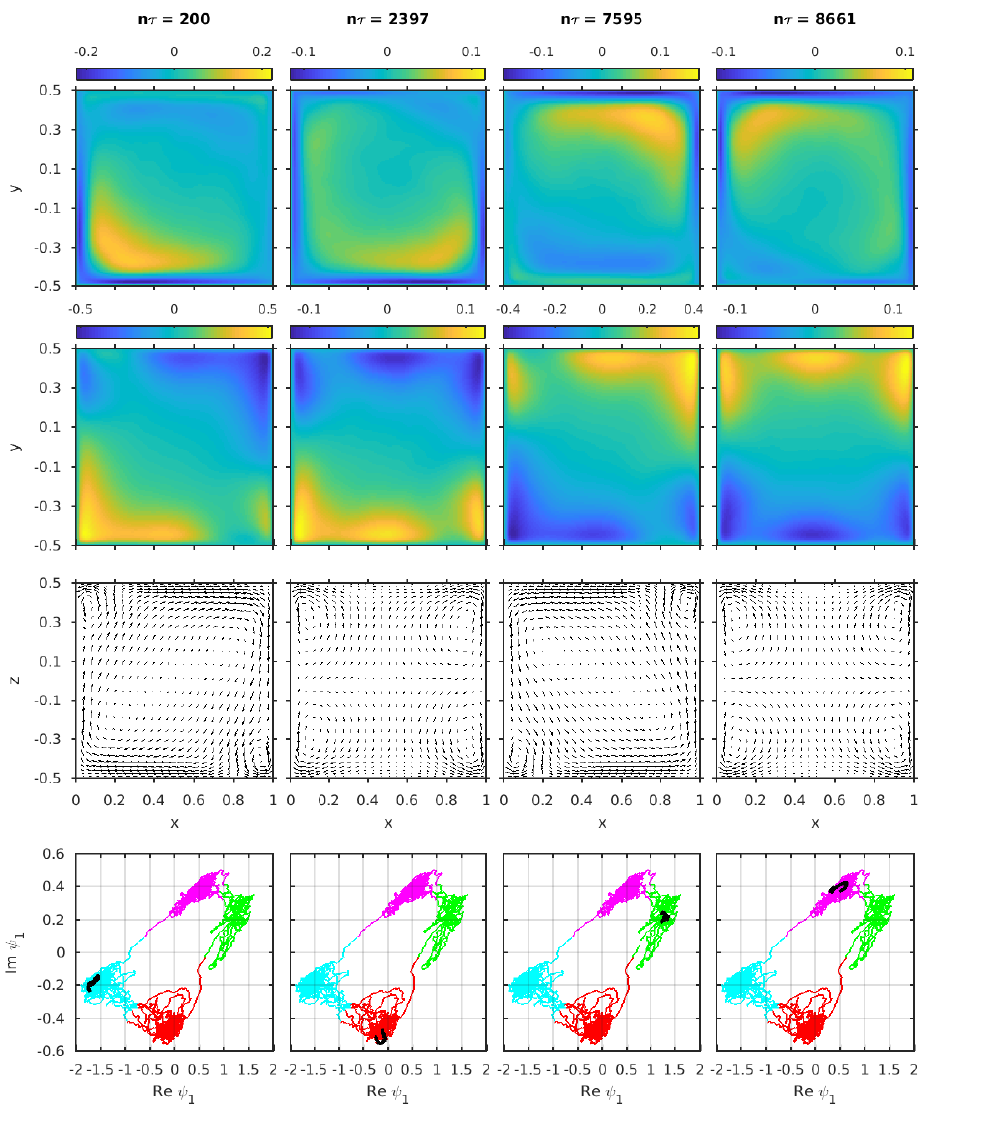}
    \caption{\label{Fig6}Reconstructed velocity field snapshots $ \hat{\bm u}^\text{primary}_n$ associated with the primary Koopman eigenfunctions (top three rows) and the corresponding phase space coordinates $(y_1,y_2,y_3)$ (bottom row). The reconstructed velocity fields were computed via~\eqref{eqPsiRec} and~\eqref{eqPrimaryRec}, and are displayed on four time instances $n\tau$, each associated with residence of the system state in one of the four phase space clusters. The cross-sections in the top three rows are as in figure~\ref{Fig5}. Dark solid lines in the bottom row indicate the phase space coordinates over the time interval $[ n\tau, n\tau + ( Q - 1 ) \tau ] $ employed for reconstruction. The snapshots at $ n\tau = 200,  2397, 7595$, and 8661 exhibit the four possible diagonal LSC states associated with clusters A, B, C and D, 
respectively (see figure~\ref{Fig3}).}
\end{figure}

Figure~\ref{Fig6} shows representative velocity field snapshots $ \hat{\bm u}^\text{primary}_n$, visualized at the same horizontal ($z=0.45$ and $z=0$) and vertical ($y=0$) planes as in figure~\ref{Fig5}, for each of the four clusters in the $(y_1, y_2, y_3)$ phase space. As stated above, the system can spend periods exceeding 1000 free-fall time units in each of these clusters, and displays infrequent rapid transitions between them. Based on the velocity field patterns at $z=0.45$, it is evident that clusters A, B, C, and D are associated with diagonal LSC states with impingement points at the $ \theta = 5 \pi / 4 $, $7\pi/4$, $\pi/4$, and $3\pi/4$ corners, respectively. Due to the asymmetry between the real and imaginary parts of eigenfunctions $ \hat{\bm \psi}_{1,2} $ and the corresponding projected patterns, the $ \theta = 3 \pi / 4 $ and $ 7 \pi /4 $ impingement points are somewhat less sharply resolved than their $\theta = \pi /4 $ and $ 5 \pi/ 4$ counterparts. Intriguingly, at the $ z =0 $ cross-section, the vertical velocity field components of the four clusters feature stronger activity adjacent to the sidewalls. As a result, clusters~A and~B (C and D) appear to form ``superclusters'', whose vertical velocity field patterns are approximately relatable by $ \theta = \pi $ rotations. 

LSC states with the qualitative features described above have previously been found in laboratory experiments \citep{Bai2016}, as well as in large-eddy simulations \citep{Foroozani2014,Foroozani2017}, of turbulent Rayleigh-B\'enard convection in a cubical domain, both at higher Rayleigh numbers. In addition,
the experiments were run over longer time intervals than our DNS study. Those studies found the flow to be trapped for long periods of the order of a few hundred free-fall times in one of the four diagonal flow states associated with the $ \theta = \pi / 4 $, $ 3 \pi / 4 $, $ 5 \pi / 4 $, and $7 \pi / 4$ corners, and subsequently switching to a rotationally adjacent macrostate. This behaviour was monitored in several different ways, e.g., by temperature probes at $z=0$ in the laboratory experiments \citep{Bai2016}, or by the vertical velocity measured as in our figure \ref{Fig3} \citep{Foroozani2017}. The latter study also found that during transitions between the diagonal macrostates the LSC exhibits wall-aligned states. Such states were called \emph{transition states}. In our reconstructions, we have also observed such wall-aligned states when the system transitions between clusters A and B, or C and D.

With regards to timescales, the angular frequency results in table~\ref{Tab2} indicate a timescale of $T_{1,2}=2\pi/\omega_{1,2}\approx 2\times 10^4$ for the expected time taken for the system to undergo a full cycle through all diagonal and wall-aligned LSC states, which is actually larger than our total observation interval. Even though this fact indicates that this $ T_{1,2}$ value should be interpreted with caution (and should be thought of as an order of magnitude estimate of the timescale of the LSC cycling process), it suggests that a sampling of all LSC states would require analysis of an even longer DNS, which was not feasible with our available resources. 

We also found that switching among the four stable diagonal LSC states is not equally probable in our analysis dataset. In particular, as can be seen in figure~\ref{Fig3}, the system exhibits periods where the LSC oscillates preferentially either between clusters A and B, or C and D, while transitions between A and D, or B and C, are significantly less frequent. This is noteworthy since  A$ \leftrightarrow $B, B$\leftrightarrow$C, C$\leftrightarrow$D, and D$\leftrightarrow$A should be equiprobable transitions due to the square symmetry of horizontal cross sections of the flow domain. The system thus appears to display a kind of broken ergodicity over the time. 

\subsection{Statistical ensemble analysis}
In order to analyse the asymmetry of residencies and transitions between clusters A, B, C, and D more closely, we performed an additional series of 
10 long-term simulations, each evolving for a time period of 5,000 free-fall times in the statistically steady regime. Snapshots were written out again 
each free-fall time unit. Each of the 10 cases started with a unique distribution of initial random perturbations of the equilibrium state 
with the linear temperature profile and the fluid being at rest, and was given enough time to relax into a statistically steady state. The simulations were 
conducted at a smaller grid resolution of $64^3$ points.
%------------------------------------------------------------------------------------------
\begin{figure}
\begin{center}
\includegraphics[width=5in,angle=0]{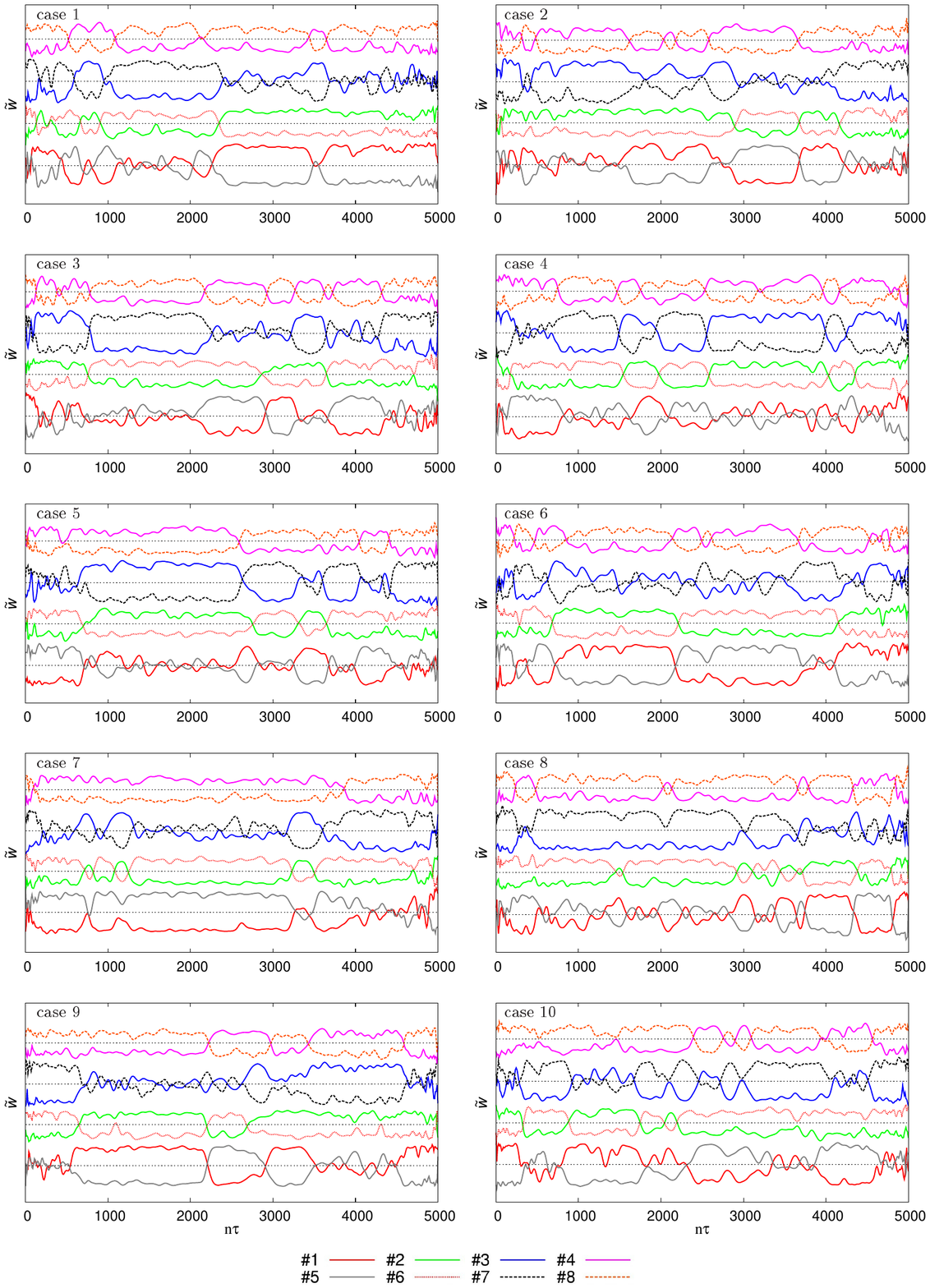}
\caption{Time series of the mid-plane $(z=0)$, filtered vertical velocity $\tilde w_{x,y}$ from~\eqref{eqUZFilt}, sampled at four points near the corners of this 
cross section (labeled \#1, \#3, \#5, \#7) and at four points near the midpoints of the sides (labeled \#2, \#4, \#6, \#8) for 10 different ensemble 
runs (shown in separate panels). Within each panel, pairs of time series are shifted vertically to 
improve visibility. Dotted lines mark $\tilde{w}_{x,y} = 0 $ axes. The numbering of the 8 probe points is as in figure \ref{Fig3}. The magnitude is again 
$|\tilde{w}_{x,y}| \le 0.5$. The time interval shown here is $ [0, 5000]$ in units of the free-fall time, which corresponds to $n\tau \in \{ 0,\dots, 5000 \} $.}
\label{Fig7}
\end{center}
\end{figure}
%-------------------------------

Figure \ref{Fig7} displays again time series of the vertical velocity resulting from an analysis similar to figure \ref{Fig3}.
This simulation ensemble exhibits a variety of different flow regimes. In several of the 10 cases, longer time intervals are detected with dominating 
diagonal flow regimes. For example, in simulations 1, 2, 6, 7, or 9, one observes an LSC roll locked in the diagonal that connects points \#1 and \#5. In cases 
1, 2, 3, 4, 8, and 9, the roll is locked in the other diagonal connecting points \#3 and \#7. Longer periods of low-amplitude oscillations at the diagonal 
corner points \#1, \#3, \#5 or \#7, as well as faster switches by 90 degrees from one diagonal state to another, are also observed in some of the cases.
To summarize this analysis: we confirm that all possible LSC states can again be found. In addition, it is even possible to detect candidates for 
LSC reversals. During such events,  all eight time series cross the zero value and the diagonal roll changes its spin staying in the same pair of corners. 
This behavior can be observed for example in case 3 at $t\approx 2900 \tau$ for corner points \#1 and \#5, or in case 4 at $t\approx 2500 \tau$ for corner 
points \#3 and \#7.
      
This ensemble analysis demonstrates that the long-term evolution may exhibit various different scenarios and switching patterns. The evolution gets 
locked in some cases in a particular regime for longer time periods. It is thus likely that our original long-term trajectory is not long enough 
to adequately sample all such regimes and transition states. This geometrically constrained convection flow thus obeys  a 
kind of broken ergodicity, as it is the case in other systems such as spin ensembles in condensed matter physics \citep{Palmer1982}. Addressing this problem in full completeness is beyond the scope of this work. We also wish to point once more to the experiments of \cite{Bai2016} and stress that they were actually run ten times longer than our DNS, namely 500,000 seconds which corresponds to about 100,000 $T_f$, all this at a Rayleigh number larger than ours by a factor of 50.

\subsection{\label{secSecondary}Secondary eigenfunctions}

\begin{figure}
\begin{center}
\includegraphics[width=5.1in,angle=0]{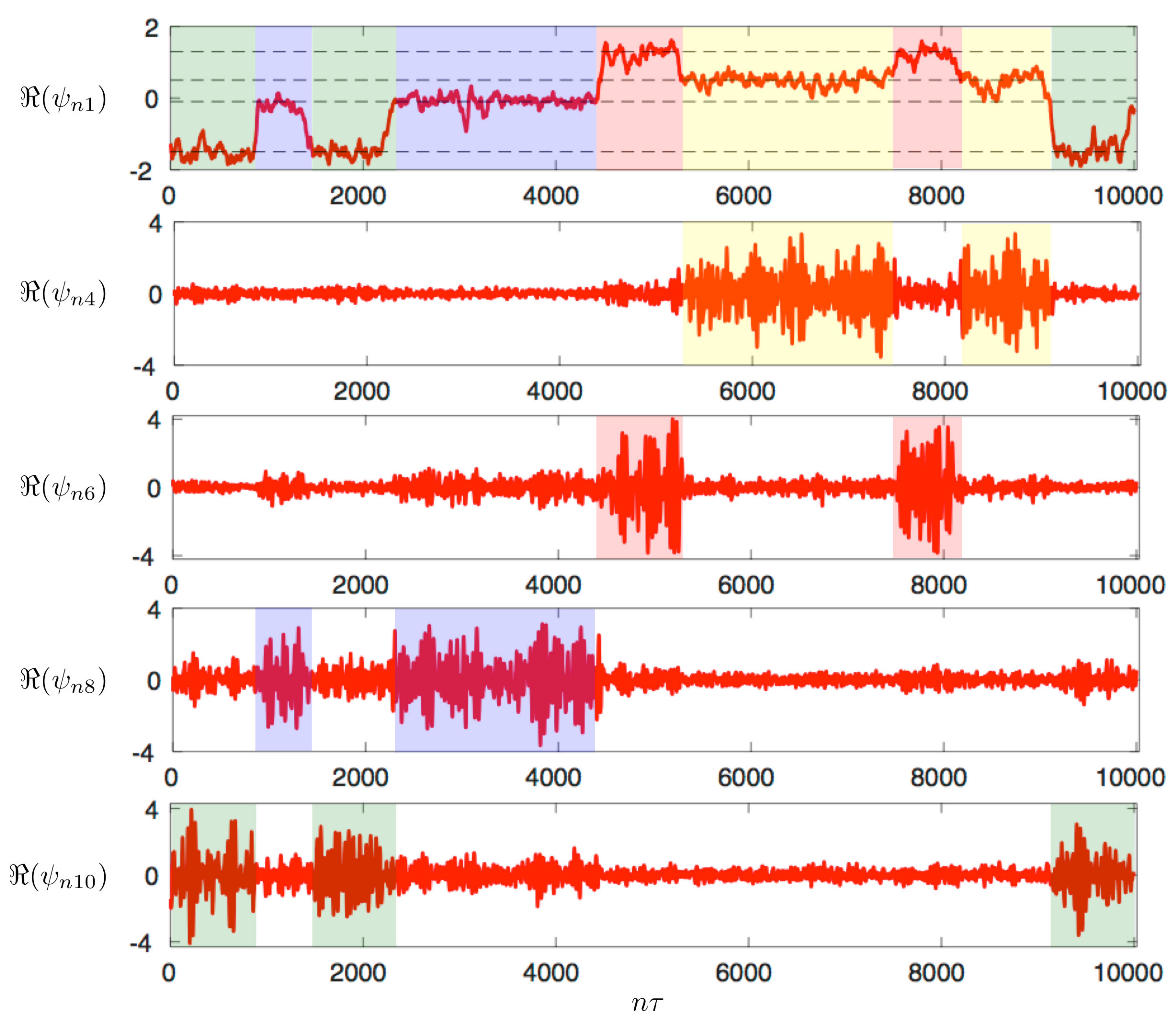}
\caption{Modulating relationships between primary and secondary Koopman eigenfunctions. We display real parts only. The top panel shows the first primary eigenfunction.  Dashed horizontal lines indicate the four clusters (diagonal macrostates) which are also seen in the central trajectory plot in figure \ref{Fig4}. Different diagonal macrostates in correspondence with figure \ref{Fig3} are indicated by different background colors (A=green, B=blue, C=red, D=yellow). The four subsequent panels below the top panel display one eigenfunction from each of the four pairs of secondary eigenfunctions. In each of the lower four panels, we mark with colored backgrounds intervals of residence in one of the macrostates that correspond to periods of enhanced fluctuations of the secondary eigenfunctions.}
\label{Fig8}
\end{center}
\end{figure}

The three primary eigenfunctions are followed by a group of eight secondary eigenfunctions, $\bm \psi_4,  \dots,  \bm \psi_{11}$, which form four complex-conjugate pairs, all of which exhibit nearly equal Koopman frequencies $ \omega_k $ as shown in table~\ref{Tab2}. Figure \ref{Fig8} displays representative eigenfunction time series from each of the four pairs. These time series are compared with the time evolution of primary eigenfunction ${\bm \psi}_1$. All four secondary eigenfunction time series display time intervals with enhanced and strongly fluctuating amplitudes which are interrupted by periods of ambient low-amplitude variations. 

The comparison with the primary mode indicates clearly that high-fluctuation intervals of each of the four pairs are synchronized with one of the four diagonal macrostates (see figure~\ref{Fig3}). For example, when the LSC flow is in cluster A (see also figure \ref{Fig4}), the complex-conjugate pair $( \bm{\psi}_{10}, 
\bm{\psi}_{11} )$ fluctuates strongly in comparison with the other  three pairs of secondary eigenfunctions. Similarly, the switch into macrostates  D, C, and B is in line with a switch to enhanced fluctuations of $(\bm{\psi}_4, \bm{\psi}_5) $,  $(\bm{\psi}_6, \bm{\psi}_7) $, and $(\bm\psi_8, \bm\psi_9)$, respectively. This  suggests that the four pairs of secondary eigenfunctions could play an important role in exchanging  kinetic energy between 
the four diagonal LSC configurations and degrees of freedom at smaller scales. If so, these patterns would be important building blocks for low-dimensional 
dynamical models of the LSC of this systems. The construction of such models is beyond the scope of this work.
%------------------------------------------------------------------------------------------
%------------------------------------------------------------------------------------------

The typical timescale $T_k=2\pi/\omega_k$ for the secondary eigenfunctions determined from their corresponding eigenfrequency is approximately 40 free-fall time units.  These quantities correspond to the time scale on which the secondary modes undergo oscillations in their active state. As with the primary eigenfunctions, we should interpret this time scale as an order of magnitude estimate, though note that in this case $ T_k $ is significantly shorter than our analysis timespan and thus has higher robustness. The synchronization to the LSC states displayed in figure \ref{Fig8} suggests that $T_k$ is a mean turnover time of a fluid parcel in the LSC roll. It is this time scale which drives corner vortices and horizontal swirls that in turn feed the LSC or cause the crossover to another macrostate. The order of magnitude of 
$T_k $ agrees well with turnover times that have been found in a Lagrangian analysis of a turbulent convection flow in a cylindrical cell of aspect ratio 1 at the same Rayleigh number \citep{Emran2010}.

\begin{figure}
\begin{center}
\includegraphics[width=\linewidth]{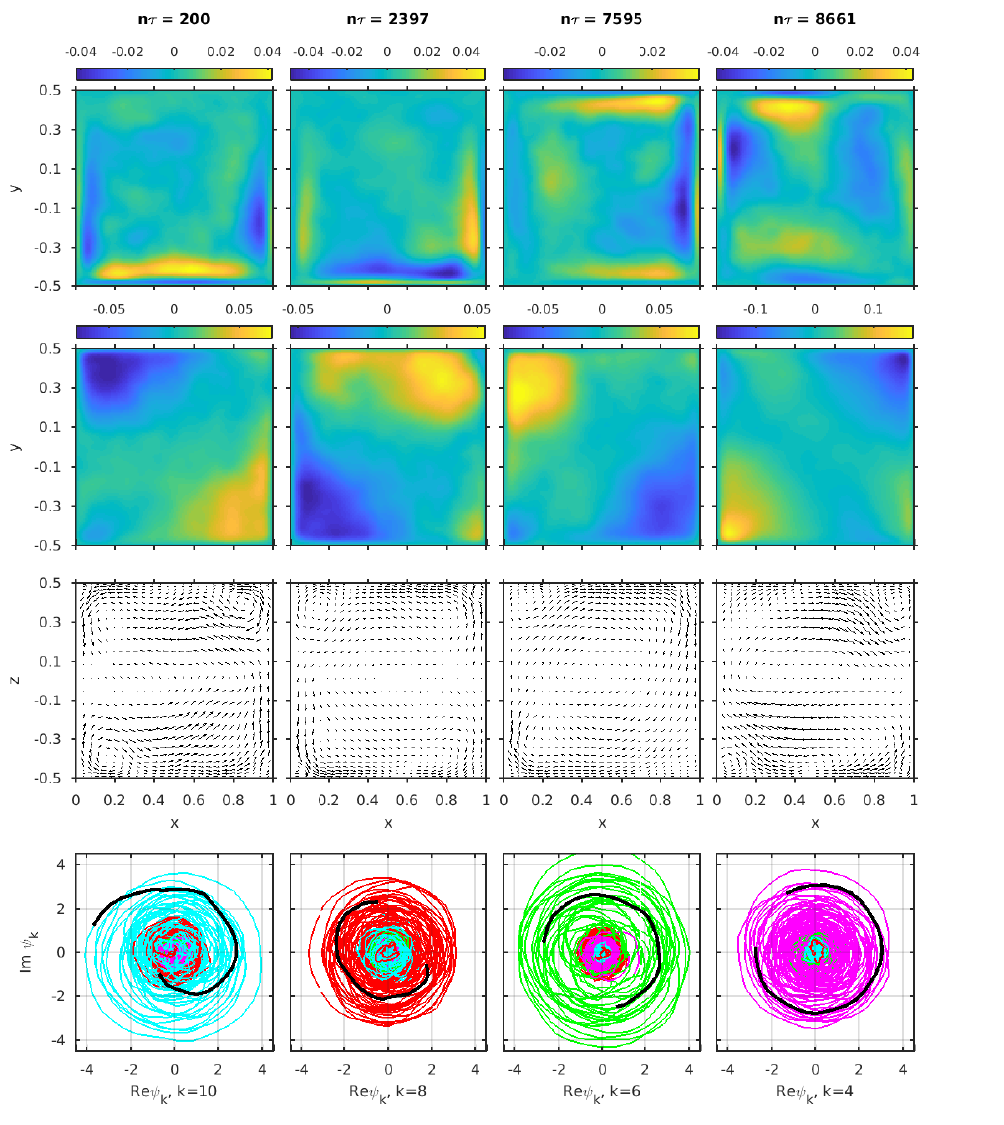}
\caption{Velocity-field reconstructions based on secondary Koopman eigenfunctions. Time instances and displayed fields in the upper three rows 
are as in figure \ref{Fig6}. In each column, we only use those eigenfunctions that fluctuate most strongly for the corresponding diagonal state. From left to right, these are: $(\bm\psi_{10}, \bm\psi_{11})$ in connection with macrostate A,  $(\bm\psi_{8}, \bm\psi_{9})$ with B,  $(\bm\psi_{6}, \bm\psi_{7})$ with C, and  $(\bm\psi_{4}, 
\bm\psi_{5})$ with D. The bottom row shows phase portraits of the secondary eigenfunctions in each column, with each time instance colored in accordance to the primary cluster affiliation as in figure~\ref{Fig3}. The dark solid line corresponds to the eigenfunction values used for reconstruction via~\eqref{eqPsiRec}, indicating the strong activity of the secondary eigenfunction pair used for reconstruction in each column.}
\label{Fig9}
\end{center}
\end{figure}

Figure \ref{Fig9} displays velocity field reconstructions based on the secondary eigenfunctions at the same time instants as data shown in figure~\ref{Fig6}. 
As in the case of the primary eigenfunctions, we display the  vertical velocity component at two horizontal cross-sections and the $(x,z)$ components in a vertical cross-section. At each time instance, we show the reconstruction from those eigenfunctions which fluctuate most strongly when the corresponding macrostate is established (see also figure~\ref{Fig8} for comparison). For example, the second column of figure \ref{Fig6} exhibits  macrostate B, and this should be considered together with 
the second column of figure \ref{Fig9} showing reconstructions based on $ \bm\psi_8$ and $\bm\psi_9$. The phase portraits in the bottom 
row of the figure confirm this relationship. That is, the eigenfunction values participating in each reconstruction (indicated by solid dark lines) are always found at larger radii where the activated secondary modes oscillate. The phase portraits in the bottom row of figure~\ref{Fig9} also demonstrate that the real and imaginary parts of the secondary eigenfunctions are to a good approximation in a 90$^\circ$ phase difference, representing a coherent amplitude-modulated oscillator.
%------------------------------------------------------------------------------------------
%-------------------------------

As shown in figure~\ref{Fig9}, the secondary eigenfunctions correspond to secondary structures in the convection flow which can appear in multiple 
configurations, including vortices that co- or counter-rotate with the LSC, corner vortices, and prominent up-or downdrafts in the midplane or close to
the plate. The vortices in the bulk and the corners can be interpreted as the flow structures that generate the oscillatory patterns in the time series of figure \ref{Fig8}. They are potentially also responsible for the eventual switch of the LSC roll out of one of the four stable macrostates into another one. Note also that the amplitudes of the primary eigenfunctions are not significantly larger (i.e., larger by more than an order of magnitude) than those of secondary eigenfunctions. This confirms that it requires a strong perturbation to drive the LSC roll out of the quasi-stable equilibrium configuration into a different one.

\subsection{\label{secSensitivity}Sensitivity analysis}
In separate calculations, we have verified that the properties of the eigenfunctions listed in table~\ref{Tab2} are robust under changes of NLSA and Koopman parameters. Among these parameters, a particularly important one is the number of delays $Q$, as it controls the ability of the NLSA eigenfunctions to span the discrete subspace $\mathcal{D}$ of the Koopman operator (see section~\ref{secNLSA}). Here, we have computed NLSA and Koopman eigenfunctions using $Q = 15 $ and $ 60 $ delays, and found that the recovered primary and secondary eigenfunction time series remain qualitatively unchanged. The frequencies $\omega_k$ of the secondary eigenfunctions typically changed by less than 5\% in these calculations. As expected from the fact that they lie close to zero (in comparison with the frequency resolution $\delta \omega$ afforded by our dataset), the frequencies of the primary eigenfunctions exhibited significantly larger, 50\% to 75\% changes, but still captured an $O(10^{-4})$ frequency scale. The primary frequencies were also more sensitive to the diffusion regularization parameter $ \zeta $ (likely for the same reasons). In particular, we found that increasing $\zeta $ to $\simeq 10^{-3}$ caused the primary eigenfunctions to become purely diffusive, with $ \omega_k $ collapsing to zero. While the LSC states discussed in section~\ref{secPrimary} are still captured by these diffusive modes, the ability to separate the two pairs of diagonal states into distinct modes is somewhat degraded (i.e., the diagonal states in figure~\ref{Fig5} become mixed). Similarly, we found that the primary eigenfunctions are more sensitive to the effects of diffusion if the finite-difference scheme is used instead of the logarithm scheme to approximate the Koopman generator $V$ (see section~\ref{secKoopEig}). As a final sensitivity test, we have examined the dependence of our results on the dimension $ \ell $ of the Galerkin approximation space, and found that satisfactory primary eigenfunctions can be obtained with $ \ell $ as low as 100. Secondary eigenfunctions capturing similar timescales as those listed in table~\ref{Tab2} can also be obtained using $ \ell \simeq 100 $, but the modulating relationships between primary and secondary eigenfunctions discussed in section~\ref{secSecondary} were found to require larger $\ell $. 
    
Overall, the observations described above highlight the fact that successful, data-driven recovery of LSC patterns in turbulent convective flows via Koopman operator techniques depends strongly on (i) the basis functions $ \bm \phi_k $ used for the Galerkin approximation (here, depending on the number of delays $Q$) (ii) 
the properties of diffusion regularization (quantified by $\zeta$), and (iii)  the approximation scheme for the generator $V$.

It should be noted that, in addition to the three points listed above, the approximation accuracy of our scheme also depends on the sampling interval $ \tau $. Due to the high computational cost of the NLSA basis calculation (which would have to be repeated if $\tau$ were to be changed), we have not performed a detailed sensitivity analysis for this parameter. In general, for a fixed number of analysis samples $N$, which may be limited by the available computational resources, one seeks a compromise between approximation accuracy for the generator (which increases with decreasing $ \tau $) and accuracy of ergodic time averaging (which increases with increasing $N\tau$). Our choice $ \tau = 1 T_f $ reflects such a compromise for patterns evolving on timescales between $ O(10 T_f) $ (secondary modes) and $ O(10^3 T_f) $ primary modes. An alternative numerical approach, lying beyond the scope of the present work, could be to maintain a $ \tau = O(1 ) $ sampling interval in the computation of the NLSA basis, but use a smaller interval $ \tau' $ (perhaps as small as a simulation timestep) to locally approximate the action of the generator on the NLSA basis functions. This would require the evaluation of NLSA eigenfunctions at states $x'_n $ sampled at times $ t'_n = n( \tau + \tau' ) $, which can be performed efficiently using out-of-sample extension techniques for Laplacian eigenfunctions \citep{CoifmanLafon06b}.      

\section{\label{secSummary}Summary}
We have performed a Koopman eigenfunction analysis to study the long-term evolution in direct numerical simulation of  three-dimensional turbulent Rayleigh-B\'{e}nard convection 
flow in a closed cubic cell. This flow is statistically inhomogeneous, and exhibits no continuous symmetry with respect to spatial coordinates. The 
large-scale circulation which builds up in such a flow has four preferential states, namely circulation rolls that fill the whole box, spin clockwise or counterclockwise, and 
are locked in diagonal corners of the cube. The four possible large-scale circulation states (two diagonals times two flow directions) are 
quasistable configurations in which the flow is found to be trapped for several hundreds of free-fall time units ($T_f$). The large-scale flow switches from one of these 
states to another via four transition states in which the circulation roll is parallel to a pair of opposite side faces of the cubic box. These two groups of 
four large-scale states can be transformed into each other by means of rotations by multiples of 90 degrees.

This behavior of the large-scale circulation  has been recovered from an analysis of the eigenfunctions of a data-driven Koopman operator governing the evolution observables under the turbulent convection flow.
These Koopman eigenfunctions were computed by means of a Petrov-Galerkin method for a regularized Koopman operator with point spectrum, in finite-dimensional Sobolev spaces. Bases of these function spaces were obtained by the eigenfunctions of a data-driven Markov operator, constructed by applying kernel algorithms in machine learning in conjunction with delay-coordinate maps of dynamical systems to a dataset of time-ordered velocity field snapshots spanning 10,000 $T_f.$  Advantages of this approach include favorable computational cost in high-dimensional data spaces, which is particularly important in the context of fully resolved three-dimensional simulation data, as well as convergence guarantees in a suitable asymptotic limit of large data.  

Our study demonstrates the applicability of this data-driven approach to a complex three-dimensional turbulent flow. Here, we focused on the leading few  
Koopman eigenfunctions which represent the primary and secondary large-scale structures in the convection cell; in particular, the large-scale circulation rolls described above, as well as corner vortices and horizontal swirls associated with secondary modes. The reconstruction of the flow by means of the three primary Koopman eigenfunctions reveals four clusters in which the system resides for periods of hundreds to thousands of free-fall time units. Although the simulation has been conducted for a very long time interval, we still observe that the 
switching between the four diagonal states is not uniformly distributed for the time interval that we were able to monitor and post-process. Rather, a 
preference to switch between two clusters (which one could call superclusters) is observed. Remarkably, our run of over ten thousand free-fall time units 
gives only two switches between these superclusters. This behaviour, which might be interpreted as a kind of broken ergodicity, was studied closer by an
ensemble analysis based on ten further runs for half the time and half the grid resolution of our main analysis run. It shows that other long-term dynamical scenarios are 
possible. Our study thus suggests that the necessary timescales to obtain an ergodic behavior should be significantly longer than 10,000 $T_f$, probably even longer than those in the laboratory experiments by \cite{Bai2016} in the same flow at a higher Rayleigh number. This task must be left as a future work, but
should be feasible since the necessary framework has been laid out here.

Several directions for future work on this subject are possible.
First, it would be interesting to explore whether the Koopman eigenfunctions recovered here from fully resolved three-dimensional velocity field data can also be recovered from sparsely sampled data, and thus from laboratory experiments. Second, the modes identified here could be employed in low-dimensional predictive models; e.g., using 
data-driven Koopman operators to evolve the modes \citep{BerryEtAl15,Giannakis17}, or related kernel analog prediction algorithms \citep{ZhaoGiannakis16}. Moreover, joint analyses of velocity and temperature field data, as well as analyses of moist convection models \citep{PauluisSchumacher10,SlawinskaEtAl14} and geophysical observations \citep{GiannakisEtAl15,SlawinskaGiannakis16} would provide a further 
extension. In these examples, convection typically takes place in large-aspect-ratio domains with $L\gg H$. The analysis could thus be used to detect 
and model large-scale patterns (or superstructures) of turbulent convection \citep[e.g.,][]{Emran2015}. Finally, with an appropriate scalable numerical implementation, the data-driven analysis presented here should be extendable to flows in the same (or similar) geometry at higher Rayleigh numbers.

\acknowledgements
DG received support from DARPA grant HR0011-16-C-0116, NSF grant DMS-1521775, ONR grant N00014-14-1-0150, and ONR YIP grant N00014-16-1-2649. The work of AK is supported by Grants no. SCHU 1410/18 and no. GRK 1567 of the Deutsche Forschungsgemeinschaft, the work of JS by the Priority Programme on Turbulent 
Superstructures which is funded by the Deutsche Forschungsgemeinschaft by Grant no. SPP 1881.
DK is supported by the Helmholtz Research Alliance ``Liquid Metal Technologies", which is funded by the Helmholtz Association and by Grant no.\ SCHU 1410/29
of the Deutsche Forschungsgemeinschaft. 
We acknowledge support with computer time by the large-scale project HIL12 of the John von Neumann Institute for Computing (NIC). 
We would like to thank Bruno Eckhardt, Najmeh Foroozani and Katepalli R.\ Sreenivasan for helpful discussions.   

\bibliographystyle{jfm}
%\bibliography{bibliography_JS}

\clearpage
\appendix

\section{}

In this appendix, we outline aspects of the numerical implementation and computational cost of our Koopman eigenfunction analysis. The main steps 
of our algorithmic framework are (1) computation of the NLSA Markov matrix $\mathsf{P}$ from~\eqref{eqPMat} and solution of the associated eigenvalue 
problem to obtain the $\bm{\phi}_k$ basis functions; (2) construction of the stiffness and mass matrices $\mathsf{L}$ and $\mathsf{B}$, respectively, 
in \eqref{eqLBMat} and solution of the Koopman eigenvalue problem; (3) reconstruction of spatiotemporal patterns associated with Koopman eigenfunctions. 
A high-level pseudocode of this procedure is displayed in table~\ref{tablePseudocode}. All NLSA and Koopman eigenfunction calculations reported in this 
paper were carried out using a Matlab code, available for download at \url{https://cims.nyu.edu/~dimitris}.

\begin{table}
    \caption{Pseudocode for data-driven Koopman eigenfunction analysis}
    \label{tablePseudocode}
    \begin{itemize}
        \item Input
            \begin{itemize}
                \item Time series $\{ \bm u_n \}_{n=-Q+1}^{N-1}$, $\bm u_n \in \mathbb{R}^{d'}$, of velocity field snapshots, sampled at $d'$ gridpoints every $\tau$ time units
                \item Number of delays $Q$
                \item Number of retained nearest neighbors $k_\text{nn}$
                \item Number of NLSA eigenfunctions $\ell \leq N - 1$ used in Galerkin approximation  
                \item Koopman regularization parameter $\zeta \geq 0 $
                \item Number of Koopman eigenfunctions $ \ell' \leq \ell $ to be computed 
            \end{itemize}
       \vspace{2ex}
        \item Output
            \begin{itemize}
                \item NLSA eigenvalues, $ \Lambda_0, \ldots, \Lambda_\ell $, with $ \Lambda_k \in [0,1] $
                \item NLSA  eigenfunctions, $ \bm \phi_1, \ldots, \bm \phi_\ell $, with  $ \bm \phi_k \in \mathbb{ R }^N $
                \item Koopman eigenvalues, $ \lambda_1, \ldots, \lambda_{\ell'} $, with $ \lambda_k \in \mathbb{C }$
                \item Koopman eigenfunctions, $ \bm \psi_1, \ldots, \bm \psi_{\ell'} $, with $ \bm \psi_k \in \mathbb{C}^N $
                \item Dirichlet energies, $\hat{\mathcal{E}}(\bm \psi_1), \ldots, \hat{\mathcal{E}}(\bm \psi_{\ell'})$, with $ \hat{\mathcal{E}}(\bm \psi_k) \geq 0 $
                \item Projected velocity fields, $\{ \hat{\bm A}_k( q\tau ) \}_{q=-Q+1}^0$, with $k \in \{ 1, \ldots, \ell' \}$
                \item Reconstructed velocity fields, $ \{ \hat{\bm u}^{(k)}_n \}_{n=0}^{N-1} $, with $k \in \{ 1, \ldots, \ell' \}$
            \end{itemize}
       \vspace{2ex}
        \item NLSA phase
            \begin{enumerate}[1.]
                \item Compute the $ N \times N $ pairwise distances $  d_Q( x_i, x_j ) $ using~\eqref{kernel}. For each $ i $, only retain the $ k_\text{nn} $ smallest values of $ d_Q( x_i, x_j ) $. 
                \item Symmetrize the retained distances by augmenting the list of retained distances $ d_Q( x_i, x_j ) $ for given $ i $ by $ d_Q( x_j, x_i ) $ if $ x_i $ is in the $ k_\text{nn} $ neighborhood of $ x_j $ (with respect to $d_Q$), but $ x_j $ is not in the $ k_\text{nn} $ neighborhood of $ x_i $. 
                \item Using the symmetrized pairwise distances, determine a value for the kernel bandwidth parameter $ \epsilon $ via the procedure described in \citet{BerryEtAl15} and \citet{Giannakis17}.
                \item Using the distance data from step~2 and the kernel bandwidth parameter from step~3, form the $ N \times N $ sparse Markov matrix $ \mathsf{P}$ via~\eqref{eqPMat}, treating all pairwise kernel values $ K_{ij} $ not associated with the retained distance data as zero.
                \item Compute the $ \ell $ largest eigenvalues $ \Lambda_k $ of $ \mathsf{P} $ and the corresponding eigenvectors $ \bm \psi_k $. 
            \end{enumerate}
            \vspace{2ex}
        \item Koopman eigenfunction phase
            \begin{enumerate}[1.]
                \item Form the $\ell \times \ell $ Koopman operator and inner product matrices $ \mathsf L $ and $\mathsf B $, respectively, using~\eqref{eqLBMat}.
                \item Compute the first $ \ell' $ generalized eigenvalues $ \lambda_1, \ldots, \lambda_{\ell'} $ and corresponding eigenvectors, $ \bm c_1, \ldots, \bm c_{\ell'} $ from~\eqref{evp}, ordered in order of decreasing real part.  
                \item Form the Koopman eigenfunctions $ \bm \psi_1, \ldots, \bm \psi_{\ell'} $ via $ \bm \psi_k = \sum_{j=1}^{\ell} c_{jk} \bm \phi_j / \eta_j $, where $c_{jk}$ is the $j$-th component of $\bm c_k$. 
                \item Normalize the solutions to unit norm, $\sum_{j=1}^\ell \lvert c_{jk} \rvert^{2} / \eta_j^2 = 1$.
            
                \item Compute the Dirichlet energies $\hat{\mathcal{E}(\bm \psi_1)}, \ldots, \hat{\mathcal{E}(\bm \psi_{\ell'})}$ from~\eqref{eqDirichlet}. 
            \end{enumerate}
            \vspace{2ex}
        \item Reconstruction phase
            \begin{enumerate}[1.]
                \item Compute the projected velocity field patterns, $\{ \hat{\bm A}_k( q \tau ) \}_{q=-Q+1}^0$, $1 \leq k \leq \ell'$, via~\eqref{eqAHat}.
                \item Using the Koopman eigenfunctions $\bm \psi_k $ and the projected patterns $\hat{\bm A}_k(q\tau)$, compute the reconstructed velocity fields  $ \{ \hat{\bm u}^{(k)}_n \}_{n=0}^{N-1} $. 
            \end{enumerate}
    \end{itemize}
\end{table}

The computational cost to evaluate the $N\times N$ pairwise distances $d_Q(x_i,x_j)$ for NLSA is $O( N^2(d+Q))$. This estimate takes into account the fact that $d_Q(x_i,x_j)$ can be evaluated without explicitly forming $ Qd $-dimensional delay-embedded snapshots (which would give rise to an $O(N^2Qd)$ cost). In particular, $d_Q(x_i,x_j)$ can be evaluated from pairwise distances $d_1(x_i,x_j)$ in $d$-dimensional snapshot space via the relation $Q d_Q^2(x_i,x_j) = \sum_{q=0}^{Q-1} d^2_1(x_{i-q},x_{j-q})$. Moreover, the calculation of $d_1(x_i,x_j)$ can be trivially parallelized by partitioning the input data into batches. By virtue of nearest-neighbor truncation, the memory cost for the pairwise distances is $O( k_\text{nn} N )$. The Markov matrix $\mathsf{P}$, which also exhibits an $O(k_\text{nn})$ memory cost, can be computed via operations on sparse arrays at an $O(k_\text{nn} N)$ computational cost. Note that the cost to compute $d_Q(x_i,x_j)$ is equivalent to that of computing a temporal covariance matrix in delay-coordinate space, used, e.g., in SSA and Hankel matrix DMD. However, unless additional sparsity-inducing steps are applied (whose efficacy may be questionable in the case of covariance matrices), the storage cost in SSA and Hankel matrix DMD is $O(N^2)$, as opposed to $O(k_\text{nn}N)$, $k_\text{nn} \ll N$, in NLSA. The favorable storage cost of NLSA becomes particularly important in applications involving large numbers of samples.

Once the Markov matrix $\mathsf{P}$ has been formed, we compute its $\ell $ largest eigenvalues, $ \Lambda_k $,  and the corresponding eigenfunctions, $ \bm \phi_k $, using Matlab's built-in iterative solver, \texttt{eigs}. The latter is based on the ARPACK library for Arnoldi methods. The eigenpairs $(\Lambda_k, \bm \phi_k)$ are then used to compute the $\ell \times \ell $ stiffness and mass matrices, $ \mathsf{L} $ and $\mathsf{B}$, respectively, employed in the Koopman eigenvalue problem, at $O(\ell^2 N)$ and $O(\ell^2)$ computation and storage cost, respectively. We solve the Koopman eigenvalue problem again using \texttt{eigs}.

\end{document}